\documentclass[aps,prd,nofootinbib,onecolumn,superscriptaddress,preprintnumbers,balancelastpage,longbibliography,nobibnotes]{revtex4-2}

\usepackage[dvipsnames]{xcolor}
\usepackage{graphicx,color,amsmath,amssymb,flushend,bm,mathrsfs,comment}
\definecolor{linkcolor}{rgb}{0.0, 0.28, 0.67}

\usepackage[
   colorlinks=true,
    urlcolor=linkcolor,
   anchorcolor=linkcolor,
    citecolor=linkcolor,
    filecolor=linkcolor,
    linkcolor=linkcolor,
    menucolor=linkcolor,
    linktocpage=true,
    pdfproducer=medialab,
    pdfa=true
]{hyperref}

\usepackage[capitalise]{cleveref}
\usepackage{tikz}
\usepackage{ulem}

\usepackage{amsmath}
\usepackage{physics}
\usepackage{simpler-wick}
\usepackage{amsfonts}
\usepackage{amssymb}
\usepackage[mathscr]{euscript}
\usepackage{setspace}
\usepackage{lipsum}
\usepackage{slashed}
\usepackage{cancel}
\usepackage{multirow}

\usetikzlibrary{calc}
\usepackage{siunitx}
\DeclareSIUnit{\year}{yr}
\DeclareSIUnit{\parsec}{pc}

\def\dbar{{\mathchar'26\mkern-12mu d}}

\newcommand{\bea}{\begin{eqnarray}\begin{aligned}}
\newcommand{\eea}{\end{aligned}\end{eqnarray}}

\newcommand{\mpl}{m_\text{pl}}

\newcommand{\Ap}{A^\prime}
\newcommand{\g}{\gamma}
\newcommand{\x}{\chi}

\newcommand{\eff}{\text{eff}}

\newcommand{\cm}{\text{cm}}

\newcommand{\Neff}{N_\text{eff}}
\newcommand{\meff}{M_\text{eff}}

\newcommand{\eps}{\epsilon}
\newcommand{\vv}{{\bf v}}
\newcommand{\xv}{{\bf x}}
\newcommand{\jv}{\boldsymbol{j}}
\newcommand{\kv}{{\bf k}}
\newcommand{\E}{\boldsymbol{E}}
\newcommand{\p}{\prime}
\newcommand{\w}{\omega}
\newcommand{\A}{\boldsymbol{A}}

\newcommand{\eV}{\text{eV}}

\newcommand{\keV}{\text{keV}}
\newcommand{\gev}{\text{GeV}}
\newcommand{\GeV}{\text{GeV}}
\newcommand{\MeV}{\text{MeV}}
\newcommand{\HI}{\text{HI}}

\newcommand{\mAp}{m_{A^\prime}}

\newcommand{\GHz}{\text{GHz}}
\newcommand{\THz}{\text{THz}}
\newcommand{\rhodm}{\rho_{_\text{DM}}}

\newcommand{\be}{\begin{equation}}
\newcommand{\ee}{\end{equation}}

\newcommand{\Sec}[1]{Sec.~\ref{sec:#1}}
\newcommand{\App}[1]{Appendix~\ref{app:#1}}
\newcommand{\Fig}[1]{Fig.~\ref{fig:#1}}
\newcommand{\Eq}[1]{Eq.~(\ref{eq:#1})}

\begin{document}

\preprint{FERMILAB-PUB-21-724-SQMS-T}

\title{Millicharged Relics Reveal Massless Dark Photons}

\author{Asher Berlin}
\email{aberlin@fnal.gov}

\affiliation{Center for Cosmology and Particle Physics, Department of Physics, New York University, New York, NY 10003, USA}
\affiliation{Theoretical Physics Department, Fermilab, P.O. Box 500, Batavia, IL 60510, USA}

\author{Jeff A. Dror}
\email{jdror1@ucsc.edu}

\affiliation{Department of Physics, University of California Santa Cruz, 1156 High St., Santa Cruz, CA 95064, USA
\\
and Santa Cruz Institute for Particle Physics, 1156 High St., Santa Cruz, CA 95064, USA}

\author{Xucheng Gan}
\email{xg767@nyu.edu}

\affiliation{Center for Cosmology and Particle Physics, Department of Physics, New York University, New York, NY 10003, USA}

\author{Joshua T. Ruderman}
\email{ruderman@nyu.edu}

\affiliation{Center for Cosmology and Particle Physics, Department of Physics, New York University, New York, NY 10003, USA}

\begin{abstract}
The detection of massless kinetically-mixed dark photons is notoriously difficult, as the effect of this mixing can be removed by a field redefinition in vacuum. In this work, we study the prospect of detecting massless dark photons in the presence of a cosmic relic directly charged under this dark electromagnetism. Such millicharged particles, in the form of dark matter or dark radiation, generate an effective dark photon mass that drives photon-to-dark photon oscillations in the early universe. We also study the prospect for such models to alleviate existing cosmological constraints on massive dark photons, enlarging the motivation for direct tests of this parameter space using precision terrestrial probes.
\end{abstract}

\maketitle

\tableofcontents

\section{Introduction}
\label{sec:introduction}

What is the underlying gauge symmetry of our Universe? Existing observations are consistent with an $ {\rm SU}(3) \times {\rm U}(1)$ structure at low energies.  However, there may be additional unbroken gauge symmetries accompanied by yet undetected massless vector bosons. 
To have escaped detection, the visible sector's contribution to the unbroken symmetry's conserved current must be absent or at least extremely suppressed.\footnote{With the inclusion of right-handed neutrinos ${\rm  U}(1)_{\text{B} - \text{L}}$ can be gauged, but the corresponding gauge coupling must be minuscule~\cite{Heeck:2014zfa,Hardy:2016kme}. The other known candidates are the non-anomalous currents associated with differences in lepton number, but these currents are broken by the observed neutrino masses even if neutrinos are Dirac fermions~\cite{Dror:2020fbh}.}
In this work, we focus on the former case such that the only direct interaction of the new massless boson with the visible sector is through its kinetic mixing with electromagnetism. Unfortunately, barring any additional interactions, such a massless ``dark photon" or $\Ap$ is completely decoupled from the visible sector since its effects are indistinguishable from a redefinition of the usual electromagnetic coupling. 

It is a natural possibility that the dark sector is non-minimal, such that the $\Ap$ is not the only low-energy degree of freedom. In fact, a massless dark photon can become physically observable if there exists additional matter content, denoted as $\x$, that is directly charged under the dark $\text{U}(1)$. In this case, the kinetic mixing between the dark and visible sectors has the well-studied effect of endowing $\x$ with a parametrically small but potentially detectable electromagnetic charge, commonly referred to as a ``millicharge."  A cosmological population of such millicharged particles (mCPs) interacts with both sectors, indirectly coupling the dark and visible photons. Such interactions can leave striking signatures in the cosmic microwave background (CMB) and are the focus of this work. 

Measurements of the CMB by the COBE~\cite{Fixsen:1996nj} and WMAP satellites~\cite{WMAP:2012fli} ushered in the era of precision cosmology, confirming that photons have free-streamed after decoupling from the thermal plasma in the hot early universe. In the process, their energy spectrum has preserved a nearly-perfect thermal blackbody, while the small anisotropies have continued to grow, giving rise to the large-scale structure that we see today. While recent efforts by, e.g., Planck~\cite{Planck:2018vyg} and ACT~\cite{ACT:2020frw} have continued to measure these anisotropies with increasing levels of precision over a handful of frequency regions, the most precise observations of the monopole spectrum were made over three decades ago by the FIRAS instrument aboard the COBE satellite~\cite{Fixsen:1996nj}, finding agreement with an ideal blackbody to within $\sim 1$\% throughout the entire frequency range consisting of 43 frequency bins. 

Measurements of the CMB spectrum have been used as a probe of a variety of new physics scenarios, including density perturbations during inflation~\cite{Hu:1994bz, Pajer:2012vz} and energy injection from decaying or annihilating relics~\cite{Hu:1993gc,Chluba:2011hw} (see, e.g., Ref.~\cite{Chluba:2019nxa} and references within). In this work, we focus on new physics that is weakly-coupled to electromagnetism that may slightly distort the CMB spectrum away from that of a blackbody. At redshifts $z \lesssim \order{10^6}$, the decoupling of photon number-changing processes means that non-standard processes that add or remove photons may lead to observable deviations that persist to the present time. The absence of such spectral distortions in the COBE/FIRAS dataset place some of the strongest bounds on models of light weakly-coupled physics beyond the Standard Model (SM)~\cite{Chluba:2019nxa}. For instance, the ionized plasma in the early universe modifies the photon's dispersion relation, which gives rise to a redshift-dependent effective photon mass spanning $\meff(z) \sim 10^{-14} \ \eV - 10^{-5} \ \eV$ during redshifts of $z \sim 0 - 10^6$, respectively. If, during its redshift evolution, the value of $\meff(z)$ approaches the mass $\mAp$ of a dark photon, the probability for $\g \leftrightarrow \Ap$ interconversion is resonantly enhanced. This has been used to place stringent upper bounds on the photon's coupling to massive dark photons~\cite{Mirizzi:2009iz,Mirizzi:2009nq,Kunze:2015noa,McDermott:2019lch,Caputo:2020bdy,Caputo:2020rnx,Garcia:2020qrp}.

An analogous effect occurs for massless dark photons in a background of mCP relics if the $\Ap$ acquires an in-medium contribution to its effective mass, $\meff^\p(z)$, which is comparable to the SM plasma's contribution to $\meff(z)$. As we show in this work, even cosmological populations of mCPs that make up a tiny subcomponent of dark matter (DM) or dark radiation (DR) can induce $\g \rightarrow  \Ap$ conversion in the early universe. In this way, the COBE/FIRAS measurement places a powerful bound on the existence of new unbroken gauge symmetries. In addition, in theories of massive dark photons, the effects of mCPs can even \emph{open up} distinct regions of the dark photon parameter space by preventing epochs of resonant conversion. This possibility implies an increased  importance of terrestrial experiments, whose sensitivity to dark photons is unaffected by the presence of such mCPs. We note that mCP-induced $\g \to \Ap$ signals were previously investigated in Ref.~\cite{Bogorad:2021uew}, which focused on ultralight mCP DM\@. We expand upon and go beyond this work by investigating a larger set of models (including heavier mCP DM, massless DR, and massive dark photons) and we improve upon such calculations by estimating the region where mCP self-interactions can significantly effect $\g \leftrightarrow \Ap$ conversion. 

We begin in \Sec{transitions} by giving a pedagogical overview of $\g \to \Ap$ transitions. In \Sec{cmb}, we discuss the imprints of such transitions on the CMB spectrum in regards to the COBE/FIRAS dataset. In Secs.~\ref{sec:DMlimit} and \ref{sec:DRlimit}, we illustrate how COBE/FIRAS excludes a large portion of the simplest parameter space for thermal millicharged DM (mCDM) and feebly-coupled millicharged DR (mCDR), respectively. Next, in \Sec{darkphoton}, we show how a cosmological population of mCDM can open up regions of parameter space for ultralight dark photons, further motivating terrestrial searches for such new long-range forces. Finally, in \Sec{conclusion} we conclude and discuss future directions. We also provide a series of appendices that contains additional calculational details.

\section{In-Medium Resonant Transitions}
\label{sec:transitions}

We begin by providing a pedagogical discussion of mCP-induced $\g \leftrightarrow \Ap$ interconversion. More complete derivations are provided in Appendices~\ref{app:fluid} and \ref{app:resonant}\@. Our starting point is the model of a massless, kinetically-mixed, dark photon, such that the interaction Lagrangian at low energies is described by~\cite{Holdom:1985ag}
\be
\label{eq:lagrangian}
\mathcal{L} \supset  \frac{\eps}{2} \, F^{\mu \nu} \, F^\p_{\mu \nu} + e \, A_\mu \, j^\mu + e^\p \, \Ap_\mu \, j^{\p \, \mu}
\supset e \, A_\mu \, (j^\mu + q_\x \, j^{\p \, \mu}) + e^\p \, \Ap_\mu \, j^{\p \, \mu}
~.
\ee
Here, $\eps \ll 1$ is the kinetic mixing parameter, $ F_{\mu \nu} ^{ (\prime)}$ the field-strengths, $j_\mu ^{ (\prime) } $ the current densities, and $e ^{ (\prime) }$ the gauge couplings, where quantities without or with a prime correspond to the $A_\mu$ or $\Ap_\mu$ field, respectively. In the second equality, we have diagonalized the kinetic mixing term up to $\order{\eps^2}$ by the field-redefinition $\Ap_\mu \to \Ap_\mu +  \eps A_\mu$. In this basis, the SM current $j^\mu$ solely couples to the visible photon, whereas the dark current $j^{\p \, \mu}$ couples both to the dark photon (with strength $e^\p$) and to the SM photon (with strength $\eps \, e^\p$). Thus, dark sector particles $\x$ directly charged under the $\Ap$ appear as though they are millicharged under normal electromagnetism with an effective charge of size $ q_\x \equiv \eps \, e^\p / e $.

Although massless dark photons are completely decoupled from the SM in vacuum, resonant $\g \leftrightarrow \Ap$ transitions are possible in a background of mCPs. This process is due to the SM electromagnetic field driving collective oscillations in the mCP plasma, which in turn generate dark electromagnetic fields (and vice-versa). This transfer of electromagnetic energy is maximally likely when the dispersion relations of the initial and final states are matched~\cite{Dubovsky:2015cca}. Although the per-particle probability for $\g \to \Ap$ is identical to that of $\Ap \to \g$, we focus strictly on signals arising from $\g \to \Ap$ oscillations in this work. This is because in most of the parameter space of interest, the SM energy density dominates over that of the dark sector, enhancing the total rate of photon-disappearance over that of photon-appearance (in \Sec{reliclimit} we explicitly comment on the parameter space in which this approximation does not hold). To calculate this conversion probability, we study the coupled equations of motion of the SM-like and dark photon-like states. Taking the momentum vector $k$ to be real, the conversion rate into dark photons is governed by the imaginary component of the frequency $\w$ of the SM-like state. As derived in \App{fluid},\footnote{The derivation in the appendix assumes a non-relativistic plasma, but comparing with the calculation done using thermal field theory~\cite{Braaten:1993jw, Raffelt:1996wa}, we conclude the result shown here applies to both relativistic and non-relativistic mCP relics.} the Fourier-transformed equations of motion for the spatial transverse and longitudinal field components are
\be
\label{eq:disp1body}
(\w^2 - k^2) \, 
\begin{pmatrix}
\tilde{\A}_T \\ \tilde{\A}_T^\p
\end{pmatrix}
\simeq 
\begin{pmatrix}
\meff^2 + \eps^2 \, \meff^{\p \, 2} & \eps \, \meff^{\p \, 2} \\
 \eps \, \meff^{\p \, 2}  & \meff^{\p \, 2}
\end{pmatrix}
\begin{pmatrix}
\tilde{\A}_T \\ \tilde{\A}_T^\p
\end{pmatrix}
~~,~~
\w^2 \, 
\begin{pmatrix}
\tilde{\A}_L \\ \tilde{\A}_L^\p
\end{pmatrix}
\simeq 
\begin{pmatrix}
\meff^2 + \eps^2 \, \meff^{\p \, 2} & \eps \, \meff^{\p \, 2} \\
 \eps \, \meff^{\p \, 2}  & \meff^{\p \, 2}
\end{pmatrix}
\begin{pmatrix}
\tilde{\A}_L \\ \tilde{\A}_L^\p
\end{pmatrix}
~,
\ee
where the Lorenz gauge condition fixes the zeroth components of the fields. $\meff$ and $\meff^\p$ denote effective SM and dark plasma masses, respectively,  and take the form,
\begin{equation} 
\label{eq:omega1body}
\meff^2 \equiv \frac{\w_{p, e}^2}{1 + i \Gamma/\w} - \frac{\w_{p, \HI}^2}{E _0 ^2 /\w^2 - 1}
\quad,\quad
\meff^{\p \, 2} \equiv \w_p^{\p \, 2} \times
\begin{cases}
(1+2i\Gamma^\p/\w)^{-1} & \text{(non-relativistic mCPs)}
\\
3/2 & \text{(relativistic mCPs, $\Gamma^\p \ll \w$)}
~.
\end{cases}
\end{equation} 
In \Eq{omega1body}, the three plasma frequencies, $ \omega _{ p, e} $, $ \omega _{ p, {\rm HI}} $, and $ \omega _p^\p$ are induced by free electrons, electrons bound in neutral hydrogen, and (free) mCPs, respectively, where in general the visible or dark sector plasma frequency is given by
\begin{equation} 
\w_p^{(\p) \, 2} \equiv 4 \pi \alpha^{(\p)} \, \sum_i Q_i^2 \, n_i \, \Big\langle \frac{v_i - v_i^3 / 3}{p_i} \Big\rangle
~.
\label{eq:wpGen}
\end{equation} 
The sum in \Eq{wpGen} is over all species $i$ in the visible or dark sector, with charge $Q_i$, number density $n_i$, velocity $v_i$, and momentum $p_i$, and the brackets correspond to an average over each species' phase-space~\cite{Raffelt:1996wa}. The energy scale $E_0 \simeq  \alpha ^2 m _e  / 2$ is the characteristic binding energy of neutral hydrogen. The momentum exchange rates, $\Gamma$ and $\Gamma^\prime$, result from Coulomb Scattering, $e^{-} p  \rightarrow e^{-} p$ and $\chi^{+} \chi^{-} \rightarrow \chi^{+} \chi^{-}$, in the SM and dark plasmas, respectively, and suppress the effective plasma masses if they are large compared to $\omega$. The rate $\Gamma$ for collisions between SM particles can safely be ignored, whereas $\Gamma^\p$ arises from mCP self-interactions due to $\Ap$ exchange and suppresses the dark plasma contribution to the effective $\Ap$ mass if $\Gamma^\p \gg \w$. Parametrically, we expect that for non-relativistic mCPs the rate is related to the number density and velocity by $\Gamma^\p \sim \alpha^{\p \, 2} n_\x / (m_\x^2 v_\x^3)$, while for relativistic mCPs it's related to the dark temperature $T _\x$ by $\Gamma^\p \sim \alpha^{\p \, 2} T_\x$. Since $\Gamma^\p$ is enhanced by a small mCP velocity, it can be substantial in the non-relativistic limit. We come back to this point in \Sec{DMlimit}\@. For relativistic mCPs, our interest will be in dark sectors that are weakly self-coupled and less dense than the photon bath, such that $ \Gamma^\p \ll \w$ and can be neglected (this was implicitly assumed in \Eq{omega1body} in the second line of the expression for $\meff^{\p \, 2}$).

Diagonalizing \Eq{disp1body} in the $\eps \ll 1$ limit, the dispersion relation of the SM-like eigenstate is
\be
\label{eq:disp}
\begin{array}{rlc} 
\omega ^2 - k ^2 &  \simeq  \meff^2 +  \Pi_{\g \to \Ap}  & ({\rm transverse}) \\ 
\omega ^2   & \simeq  \meff^2 +  \Pi_{\g \to \Ap}  & ( {\rm longitudinal} ) 
\end{array} \,,\qquad 
\Pi_{\g \to \Ap} \equiv \eps^2\frac{ \, \meff^{\p \, 2}\, \meff^2}{\meff^2 - \meff^{\p \, 2}}
~.
\ee
We focus solely on the second term $ \Pi_{\g \to \Ap}$ on the RHS of the dispersion relation above since it is responsible for the $\g \to \Ap$ transition. Decomposing $\omega $ and $ \Pi_{\g \to \Ap}$ into their real and imaginary components, we can solve \Eq{disp} for the real and imaginary components of $ \omega $. In the limit that $k \gg \meff , \meff^\p$ (which is the one of interest to us), the physical solution of the imaginary component of $\w$ is 
\begin{equation} 
\text{Im} (\w) \simeq 
- \left| \frac{\text{Im} ( \Pi_{\g \to \Ap})}{2 \, \text{Re}(\w)}  \right|  
\end{equation} 
for transverse or longitudinal modes.\footnote{The absolute value sign arises because we consider only the solution which depletes the SM-like state (populates the dark photon state).} From this point forward, we shorten our notation and $\w$ should be taken to mean the real part $\text{Re}(\w)$ unless specified otherwise. As in, e.g., Refs.~\cite{Weldon:1983jn,Redondo:2013lna,Hardy:2016kme}, this can be used to determine the photon phase space density $f_\g$, which evolves according to 
\begin{equation} 
\label{eq:rate1}
\frac{\partial f_\g}{\partial t} \simeq 2 f_\g \, \text{Im}(\w) \simeq - f_\g ~  \left| \frac{ \text{Im} ( \Pi_{\g \to \Ap}) }{ \omega } \right|  
~.
\end{equation} 
To evaluate \Eq{rate1}, we decompose $\meff^2$ and $\meff^{\p \, 2}$ into their real and imaginary components\footnote{Using \Eq{omega1body} we find,
$$
\text{Re}(\meff^2) = \frac{\w_{p,e}^2}{1 + (\Gamma / \w)^2} - \frac{\w_{p, \HI}^2}{E_0^2/\w^2 - 1}
~~,~~
\text{Im}(\meff^2) = - \, \frac{\w_{p,e}^2 \, (\Gamma/\w)}{1 + (\Gamma / \w)^2}
~.
$$ 
For the dark plasma, these depend on the nature of mCPs. E.g., for a non-relativistic mCP plasma,
$$
\text{Re}(\meff^{\p \, 2}) = \frac{\w_p^{\p \, 2}}{1 + (2 \Gamma^\p / \w)^2} ~~,~~
\text{Im}(\meff^{\p \, 2}) =  - \, \frac{\w_{p}^{\prime 2} \, (2\Gamma^\p/\w)}{1 + (2\Gamma^\p / \w)^2} 
~.
$$
\label{Pifootnote}
} and plug them into \Eq{disp},
which gives
\begin{equation}
\label{eq:rate2}
\frac{\partial f_\g}{\partial t} \simeq - \frac{\pi \eps^2 \, f_\g}{\w} ~ \text{Re}(\meff^2)^2 ~ \delta \Big(\text{Re}(\meff^2) - \text{Re}(\meff^{\p \, 2})\Big)
~,
\end{equation}
where we approximated $\text{Re}(\meff^2) \simeq \text{Re}(\meff^{\p \, 2})$ (i.e., near resonance) and used the narrow width approximation assuming $\text{Im}(\meff^2) , \text{Im}(\meff^{\p \, 2}) \ll \text{Re}(\meff^2) , \text{Re}(\meff^{\p \, 2})$. \Eq{rate2} is the main result of this section, which we will use to calculate the rate for $\g \to \Ap$ transitions in the early universe. Since we take the electron and mCP fluids to be weakly coupled, this transition rate is resonantly enhanced when $\w_p^2 \sim \w_p^{\p \, 2}$, where
\be
\label{eq:wpSM}
\w_p^2 \simeq \w_{p,e}^2 - (\w / E_0)^2 \, \w_{p, \HI}^2
\ee
is the approximate total contribution to the SM plasma mass. From this point forward, we abbreviate our notation such that $\meff^2$ and $\meff^{\p \, 2}$ should be taken to mean the real parts $\text{Re}(\meff^2)$ and $\text{Re}(\meff^{\p \, 2})$, unless specified otherwise. We also note that in our calculation we ignore fluctuations in the SM and mCP densities, leaving the investigation of such inhomogeneities to future work.

\section{Imprints on the CMB}
\label{sec:cmb}
Armed with the expression for the photon to dark photon conversion rate in \Eq{rate2}, we now discuss the imprint of $\g \to \Ap$ transitions on the CMB spectrum and how  non-observation of such signals bounds the strength of these processes. To begin, it is helpful to rewrite \Eq{rate2} in a form that is more applicable to signals arising in the early universe. This can be done by changing variables from time $t$ to cosmological redshift $z$ and then integrating over $z$, which gives the fractional change to the CMB spectrum observed at a frequency $\w_0 = \w / (1+z)$ today
\be
\label{eq:rate3}
\frac{\Delta f_\g (\w_0)}{f_\g (\w_0)} \simeq - \frac{\pi \eps^2}{\w_0} \sum_{z_\text{res}} \, \frac{\meff^{\p \, 2}}{  (1+z)^2 H }~ \, \bigg| \frac{d \log{\w_p^2}}{dz} - \frac{d \log{\meff^{\p \, 2}}}{dz} \bigg|^{-1} ~ \Bigg|_{z = z_\text{res}}
~,
\ee
where $H$ is the Hubble parameter and the sum is over all such redshifts $z_\text{res}$ at which the resonance $\w_p^2 \simeq \meff^{\p \, 2}$ occurs (we remind the reader that compared to the discussion above, we have shortened our notation such that only the real part is included in $\meff^{\p \, 2}$). 

Note that replacing $\meff^{\p \, 2} \to \mAp^2$ in \Eq{rate3} recovers the familiar expression typically used to derive $\gamma \to \Ap$ transitions for  massive dark photons in the absence of mCPs~\cite{Mirizzi:2009iz,Mirizzi:2009nq,Kunze:2015noa,McDermott:2019lch,Caputo:2020bdy,Caputo:2020rnx,Garcia:2020qrp}. More generally, when $\mAp \gtrsim \meff^\p$, one should replace $\meff^{\p \, 2} \to \meff^{\p \, 2} + \mAp^2$. In this sense, at early times when the density of mCPs is potentially sizeable, $\meff^\p$ plays the role of an effective redshift-dependent dark photon mass. In our analysis below, we use \Eq{rate3} to determine limits or projections from existing and future measurements of the CMB spectrum. 

In general, multiple resonances can contribute to the sum in \Eq{rate3}. In the absence of mCPs, as studied previously in, e.g., Refs.~\cite{Mirizzi:2009iz,Mirizzi:2009nq,Kunze:2015noa,McDermott:2019lch,Caputo:2020bdy,Caputo:2020rnx,Garcia:2020qrp}, $\meff^{\p \, 2} \to \mAp^2$ implies that the smallest $z_\text{res}$ typically dominates since $(1+z)^2 H$ is a rapidly growing function of redshift. In contrast, when the $\Ap$ effective mass is dominated by the mCP contribution, $\meff^{\p \, 2} \propto \w_p^{\p \, 2} (z)$ also scales rapidly with redshift due to the enhancement  of the mCP density at earlier times. For example, we find that for adiabatic transitions induced by non-relativistic mCPs during matter domination, the sum of \Eq{rate3} is dominated by the largest $z_\text{res}$.

In the early universe, the thermal blackbody distribution of photons is maintained by their tight coupling to the SM plasma such that any deviation away from equilibrium at high redshift is quickly erased by rapid collisions.  However, the decoupling of photon number-changing reactions (in the form of double Compton scattering) at redshifts of $z \lesssim 2 \times 10^6$ implies that late-time deviations away from equilibrium can persist to the present time (see, e.g., Ref.~\cite{Chluba:2011hw}). The best existing bound on such spectral distortions of the CMB arises from previous measurements by the FIRAS spectrometer aboard the COBE satellite. The CMB blackbody as observed by COBE is presented at the level of observed intensity $I _{ {\rm obs}}$, or power per unit frequency per unit area per solid angle. Approximating the unperturbed SM photon intensity as a perfect blackbody at temperature $T_0 \simeq 2.7~{\rm K} $, the perturbed intensity $I(\w_0, T_0)$ as observed today is
\be
I(\w_0, T_0) \simeq   \frac{ \w_0^3 / 2 \pi^2 }{  e^{\w_0 / T_0} - 1}\, \bigg( 1 + \frac{\Delta f_\g (\w_0)}{f_\g (\w_0)} \bigg)
~.
\ee
For a particular mCP population, we calculate the predicted intensity $I(\w_0, T_0)$ and compare to the observed intensity $I_\text{obs} (\w_0)$ from COBE~\cite{Fixsen:1996nj} by means of the  $\x^2$ test statistic 
\be
\label{eq:chi2}
\x^2 = \sum_i \left( \frac{I_\text{obs} (\w_i) - I ( \w_i, T_0 )}{\sigma  (\w_i)} \right)^2
~,
\ee
where the sum over $i$ corresponds to the set of measured frequency bins $\w_i$ with corresponding intensity uncertainty $\sigma (\w_i)$. Note that \Eq{chi2} is an approximation, as a more exact treatment would involve a convolution of the instrumental response function over the frequency range of each individual bin. We have verified that \Eq{chi2} agrees with performing a full convolution to within $\mathcal{O}(10\%)$. 

 In determining limits on models of mCP relics, we allow the temperature $T _0$ to be a free parameter determined by minimizing the $\x^2$. This is necessary since the CMB monopole is altered in the presence of an electromagnetically-coupled dark sector. We will also illustrate the projected sensitivity of a more precise measurement of the CMB spectrum, as can be achieved by the proposed PIXIE satellite~\cite{Fixsen:1996nj}. Following the discussion in Ref.~\cite{Kunze:2015noa}, we construct a $\x^2$ for PIXIE by assuming 400 angular frequency bins of resolution $2 \pi \times 15 \ \text{GHz}$ spanning the range $2 \pi \times 30 \ \GHz - 2 \pi \times 6000 \ \THz$ with an intensity resolution of $\sigma (\w_i) \simeq 5 \times 10^{-26} \ \text{W} / \text{Hz} / \text{m}^{2} / \text{sr}$. Compared to the FIRAS instrument, this corresponds to an $\order{10}$ increase in the number of frequency bins and an $\order{10^3} - \order{10^5}$ enhancement in the intensity resolution. 

\begin{figure}[t]
\centering
\begin{tikzpicture} 
\node at (-7.5,0) {\includegraphics[width=0.479\columnwidth]{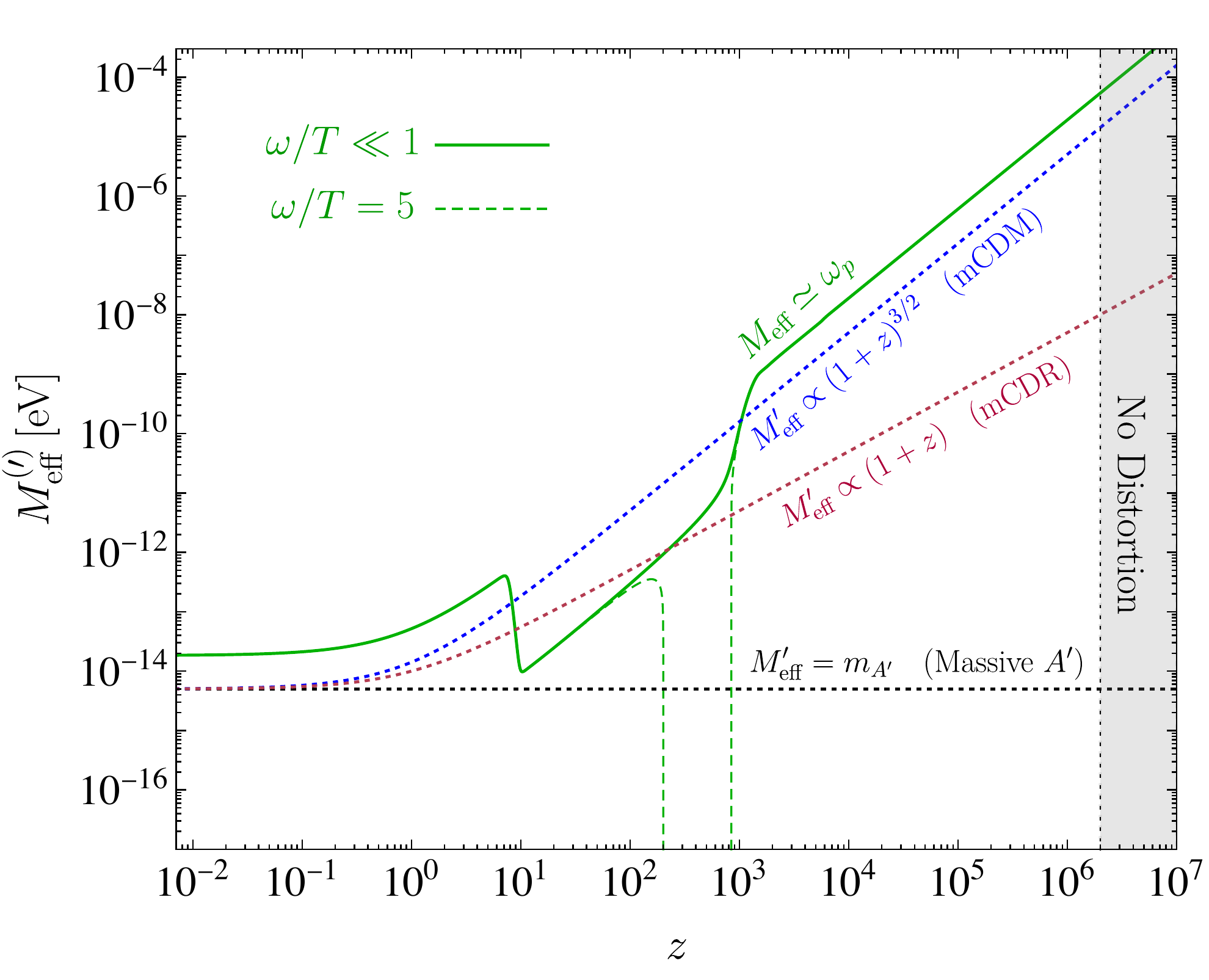}};
\node at (1.25,0.035) {\includegraphics[width=0.447\columnwidth]{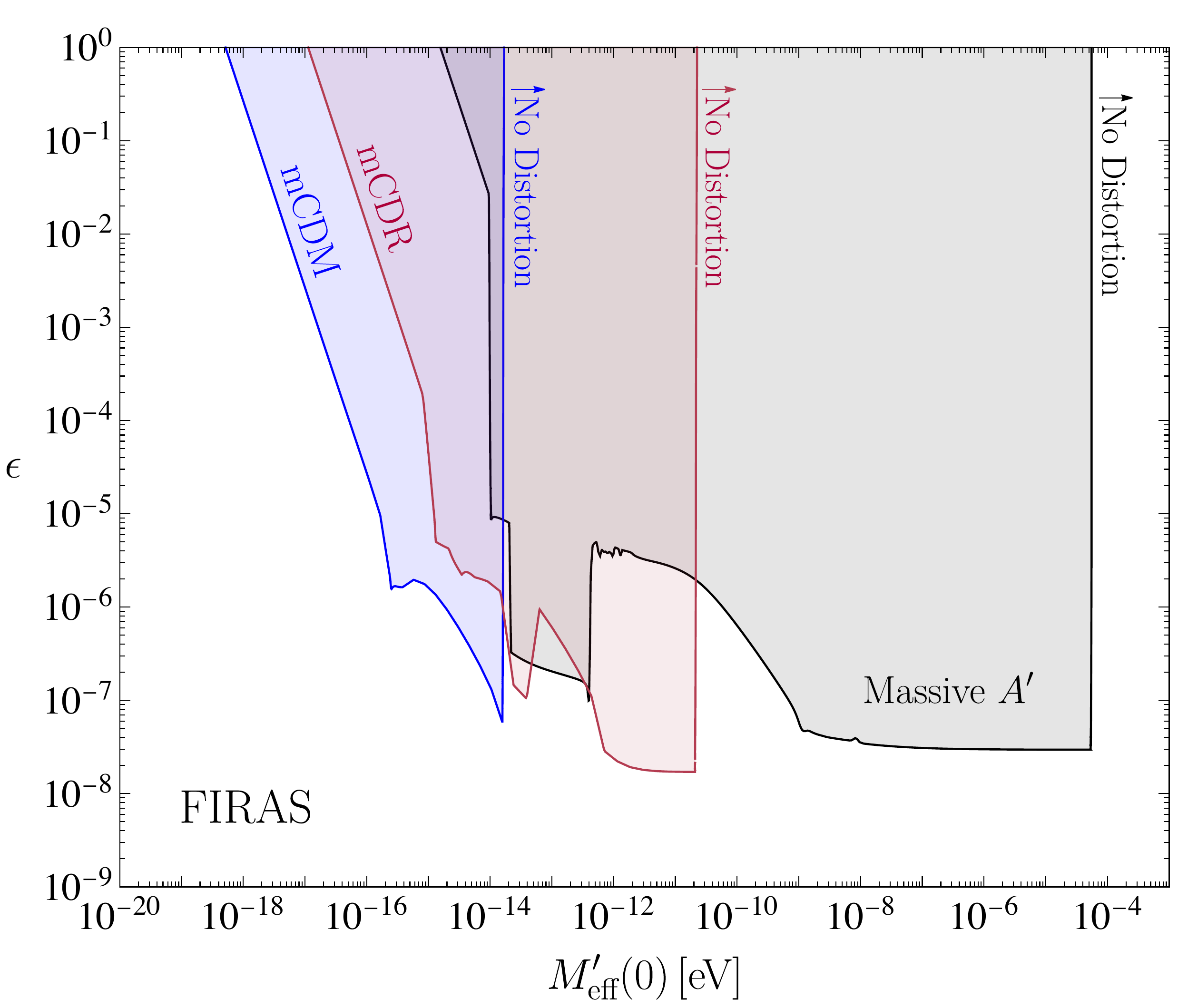}};
\end{tikzpicture}
\vspace{0.3cm}
\caption{{\bf Left}: The redshift evolution of the effective Standard Model photon mass $\meff \simeq \w_p$ for two different choices of the frequency to temperature ratio, $\w / T \ll 1$ (solid green) and $\w / T = 5$ (dashed green), as well as the effective dark photon mass $\meff^\p$ for a massless dark photon interacting with millicharged dark matter (dotted blue) or dark radiation (dotted red). For comparison we also show $\meff^\p = \mAp$, corresponding to a massive dark photon absent of millicharged relics (dotted black). Level crossings corresponding to points at which $\meff^\p \simeq \w_p$ induce $\g \to \Ap$ transitions that distort the CMB spectrum.  Transitions at redshift $z \gtrsim 2 \times 10^6$ (shaded black region) are quickly erased by rapid photon number-changing reactions. {\bf Right}: Constraints derived from COBE/FIRAS measurements of the CMB spectrum in the parameter space spanned by the dark photon kinetic mixing parameter $\eps$ and the current $z = 0$ value of the effective dark photon mass $\meff^\p (0)$ for models of massless dark photons and millicharged dark matter (blue) or dark radiation (red). For comparison we also show the bounds on models of massive dark photons absent of millicharged relics (black), which agrees with the previous results of, e.g., Refs.~\cite{Mirizzi:2009iz,Mirizzi:2009nq,Kunze:2015noa}.}
\label{fig:zevolution}
\end{figure}

The large density of mCP relics in the early universe can lead to resonant production of dark photons if $\meff^{\p \, 2} (z_\text{res}) \simeq \w_p^2 (z_\text{res})$ at redshift $z_\text{res}$. In this work, we focus on mCPs that constitute a small fraction of the total dark matter (mCDM) or radiation (mCDR) energy densities. For a fixed comoving number of weakly-coupled mCPs, the redshift dependence of the dark plasma frequency squared scales as the number density of mCDM or as the average momentum squared of mCDR\@. As a result, we can characterize the effect of mCPs as,
\be
\label{eq:meffgen}
\meff^\p (z)  \propto (1+z)^n  ~~,~~
n =
\begin{cases}
3/2 & (\text{mCDM})
\\
1 & (\text{mCDR})
\end{cases}
~.
\ee
 In the left panel of \Fig{zevolution}, we show the redshift evolution of $\meff \simeq \w_p$ (green lines) and $\meff^\p$ (dotted blue and red lines). The redshift evolution power index $n$ defines the type of mCP dark sector. If no mCPs are present, then for a massive dark photon $\meff \simeq \mAp$ is independent of redshift (dotted black line). For the SM contribution, we show $\w_p$ for two representative values of the frequency $\w$ (solid and dashed green lines). For larger values of $\w$, $\w_p ^2 $ is driven to negative values for particular redshift regions by the neutral hydrogen density (see \Eq{wpSM} and footnote~\ref{Pifootnote}). As mentioned above, for redshifts $z \gtrsim 2 \times 10^{6}$, double Compton scattering is efficient and thus erases any deviations of the SM photon spectrum away from that of an ideal blackbody. This region is shaded black in the left panel of \Fig{zevolution}.

The redshift dependence of $\w_p$ and $\meff^\p$ can be used to determine the resulting modifications to the CMB spectrum, as detailed above. Note that $\meff^\p(z)$ is specified by its current $z  =0$ value $\meff^\p (0)$ and its redshift scaling $n$. In this sense, $\meff^\p(0)$ is a useful effective parameter. For the remainder of this section, we therefore choose to present limits on the dark sector model space in terms of $\eps$ and $\meff^\p (0)$ for theories of massless dark photons ($n = 0$), mCDM ($n = 3/2$), and mCDR ($n = 1$). In the next section, we will illustrate more explicitly how such limits are placed within the detailed model space of mCP relics, i.e., how specific models and cosmologies map onto $\meff(0)$ and $n$. The limits as derived from FIRAS are shown in the right panel of \Fig{zevolution} in the parameter space spanned by the dark photon kinetic mixing parameter $\eps$ and $\meff^\p (0)$ for models of massless dark photons and mCDM (blue) or mCDR (red). For simplicity, we have ignored the mCP self-scattering rate $\Gamma^\p$ such that $\meff^\p \sim \w_p^\p$ (we discuss this in more detail in \Sec{reliclimit}). For comparison we also depict the bounds on models of massive dark photons absent of mCPs (black), as calculated previously~\cite{Mirizzi:2009iz,Mirizzi:2009nq,Kunze:2015noa}. In deriving these limits, we have evaluated the $\x^2$ test statistic from \Eq{chi2} after marginalizing over the current CMB temperature $T_0$ for a particular choice of model parameters. In this three-dimensional parameter space  spanned by $\meff^\p (0)$, $\eps$, and $T_0$, the excluded regions are shown for $\x^2 - \x_\text{min}^2 = 7.82$, corresponding to a 95\% confidence limit, where $\x_\text{min}^2$ is the minimum value of $\x^2$ within the entire parameter space shown. 

FIRAS is sensitive to values of the kinetic mixing as small as $\eps \sim 10^{-7}$ in each of the models highlighted in \Fig{zevolution}. In models of mCDM, both the dark sector and SM plasma masses scale as $\meff^\p \propto \w_p \propto z^{3/2}$ at high redshift. As a result, for sufficiently large values of $\meff^\p (0)$, the condition $\meff^\prime \simeq \w_p$ is never satisfied and hence resonant $\Ap$ production does not occur. This corresponds to the sharp edge near $\meff^\p (0) \sim 10^{-14} \ \eV$ in the blue shaded region in the right panel of \Fig{zevolution}. On the other hand, for mCDR $\meff^\p \propto (1+z)$ implies that resonances always occur for sufficiently large values of $\meff^\p (0)$. However, since any distortions induced by $\g \to \Ap$ transitions that occur before $z \simeq 2 \times 10^6$ are quickly suppressed by double Compton scattering, the red shaded region in the right panel of \Fig{zevolution} does not extend well past $\meff^\p (0) \sim 10^{-11} \ \eV$\@.

From the left panel of \Fig{zevolution}, we see that for sufficiently small effective dark photon masses, $\meff^\p (0) \ll 10^{-14} \ \eV$, resonant transitions are possible only for $\w / T \gtrsim \text{few}$, in which case neutral hydrogen (corresponding to the second term of \Eq{wpSM}) is able to suppress the visible plasma frequency down to $\w_p \simeq \meff^\p$ in a narrow range of redshift. In this small $\meff^\p$ limit, we find that the conversion probability is well approximated by
\be
\label{eq:P_AToAp_SmallmAp}
\lim_{\meff^\p (0) \ll 10^{-14} \ \eV}  \frac{\Delta f_\g(\w_0)}{f_\g(\w_0)} = -\frac{\pi \, \epsilon^2 \, \meff^{\p \, 4} (0)}{\left( 2 - \delta \right) H_0 \, \Omega_{m}^{1/2}}  \frac{\mathcal{C}_2^{\frac{11/2 - 4n + \delta}{2-\delta}}}{\mathcal{C}_1^{\frac{15/2-4n}{2-\delta}}} ~ \w_0^{\frac{9-8n+3\delta}{2-\delta}}
~,
\ee
where $n$ is defined in \Eq{meffgen}, $\mathcal{C}_1 \equiv 3.5 \times 10^{-32} \ \eV^2$, $\mathcal{C}_2 \equiv 1.9 \times 10^{-30}$, $\delta \equiv 0.19$, $H_0$ is the present-day value of the Hubble parameter, and $\Omega_m \simeq 0.3$ is the fractional cosmological matter density. Thus, in this low mass range, the limit on kinetic mixing scales as $\epsilon \propto \meff^{\p \, -2} (0)$, as evident by the left-most part of the exclusion contours in the right panel of \Fig{zevolution}. Alternatively, for very large values of $\meff^{\p} (0)$, resonant $\g \to \Ap$ transitions occur deep into radiation domination. In this case, we find that the conversion rate is instead approximated by
\be
\label{eq:P_AToAp_LargemAp_Limit}
\lim_{\meff^\p (0) \gg 10^{-14} \ \eV}   \frac{\Delta f_\g(\w_0)}{f_\g(\w_0)} = -\frac{\epsilon^2 }{3-2n}\sqrt{ \frac{45}{4 \pi \, g_*} } \frac{\mpl  \, \meff^{2} (0)}{ \omega_0 \, T_0^2 }
~,
\ee
where $g_*$ is the number of effective relativistic degrees of freedom in the SM bath at $z_\text{res}$. \Eq{P_AToAp_LargemAp_Limit} implies that in this high mass range, the limit on kinetic mixing is independent of $\meff^{\p} (0)$, as evident by the right-most part of the exclusion contours in the right panel of \Fig{zevolution}.

\section{Constraining Millicharged Relics}
\label{sec:reliclimit}
In the previous section, we calculated limits as a function of $\eps$ and $\meff^\p (0)$ for models of mCDM and mCDR (see the right panel of \Fig{zevolution}). These bounds can be translated straightforwardly into the mCP parameter space spanned by $\eps$, the mCP mass $m_\x$, and the dark gauge coupling $e^\p$, assuming a fixed cosmological history. In this section, we investigate how these bounds apply to the simplest cosmologies of mCP dark sectors. 

\subsection{Subcomponent of Millicharged Dark Matter (mCDM)}
\label{sec:DMlimit}

Let us begin by investigating the case of a millicharged dark matter (mCDM) subcomponent. The effective dark photon mass arising from such a population of mCPs, assuming that their self-interactions are negligible ($\Gamma^\p \ll \w$), is approximately 
\be
\label{eq:wp}
\meff^\p \simeq \bigg( \frac{4 \pi \alpha^\p \, n_\x}{m_\x} \bigg)^{1/2} \sim 10^{-17} \ \eV \times (1+z)^{3/2} ~  \bigg(\frac{\alpha'}{10^{-4}}\bigg)^{1/2} \bigg(\frac{m_\x}{1 \ \gev}\bigg)^{-1} \bigg(\frac{f_\text{DM}}{10^{-2}}\bigg)^{1/2}
 ~,
\ee
where $\alpha^\p = e^{\p \, 2} / 4 \pi$ is the dark fine structure constant and in the second equality we have normalized the mCP number density $n_\x$ by the fractional contribution $f_\text{DM} = m_\x \, n_\x / \rhodm$ to the total DM energy density $\rhodm$.  In the following, we only consider $f_\text{DM} \lesssim 0.4 \times 10^{-2}$, since for such fractions the stringent CMB bounds on DM-baryon scattering do not apply ~\cite{Dubovsky:2003yn, dePutter:2018xte, Xu:2018efh,Kovetz:2018zan,Buen-Abad:2021mvc, Nguyen:2021cnb}.

\begin{figure}[t]
\centering
\includegraphics[width=0.49\columnwidth]{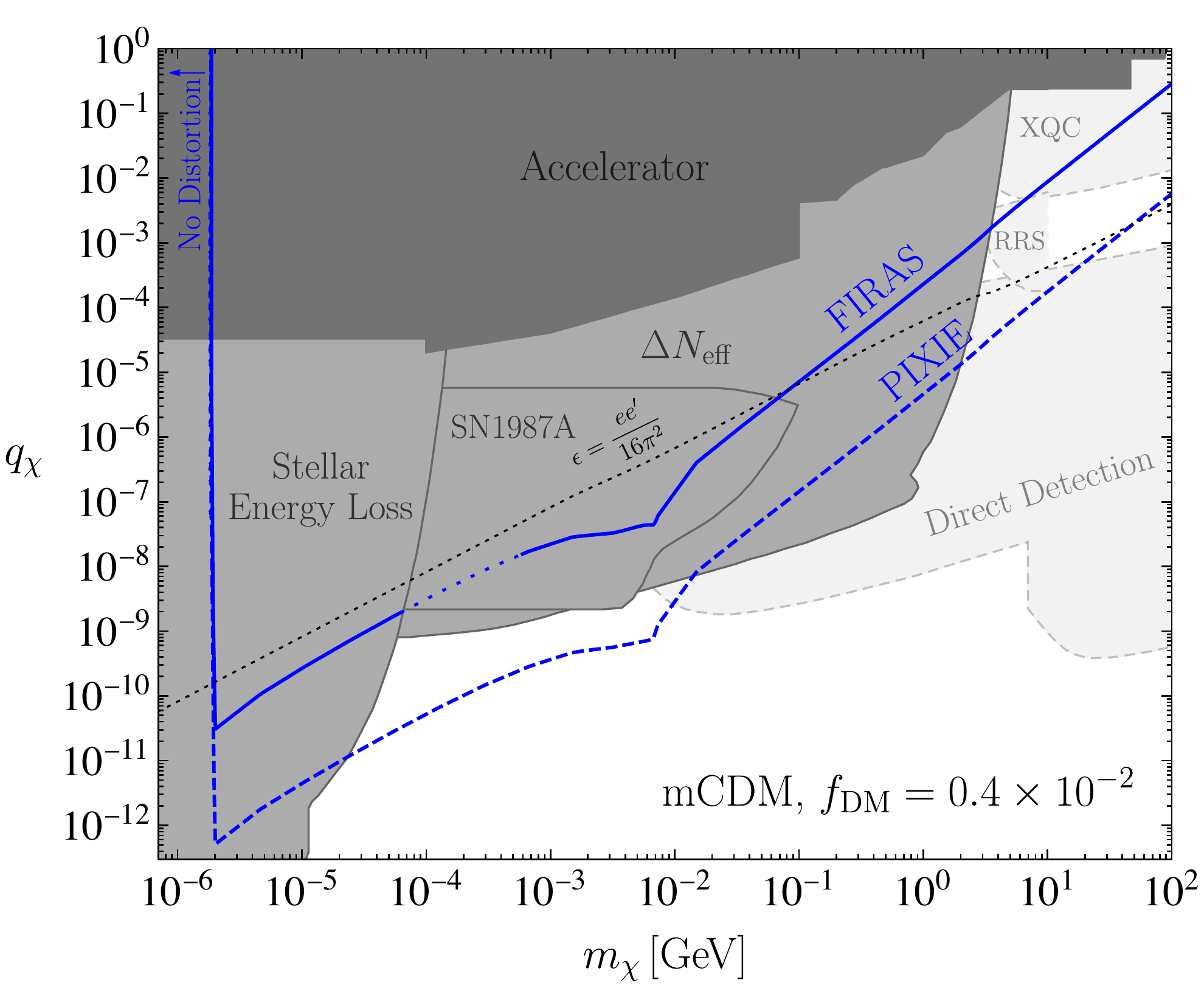}
\includegraphics[width=0.49\columnwidth]{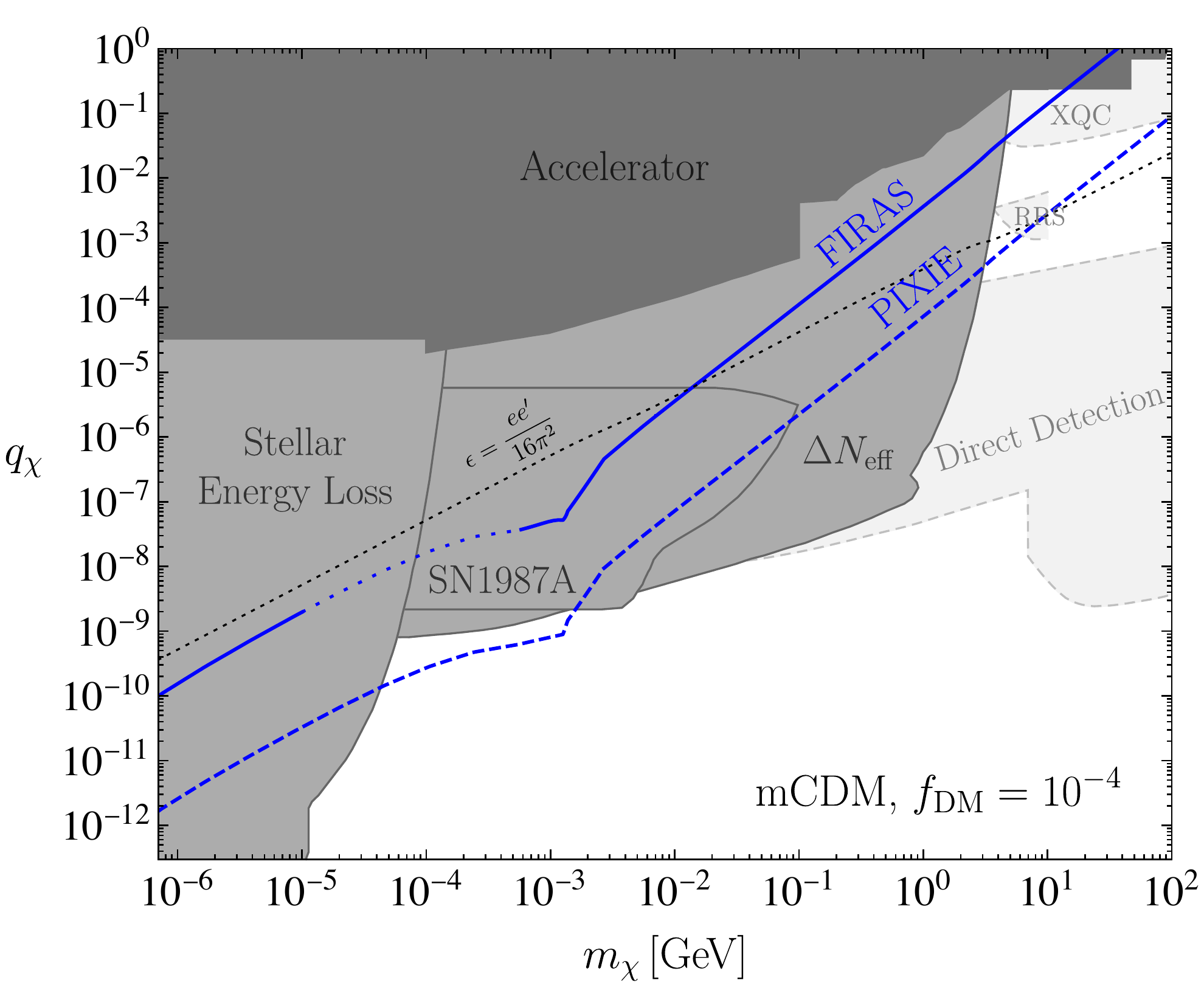}
\includegraphics[width=0.49\columnwidth]{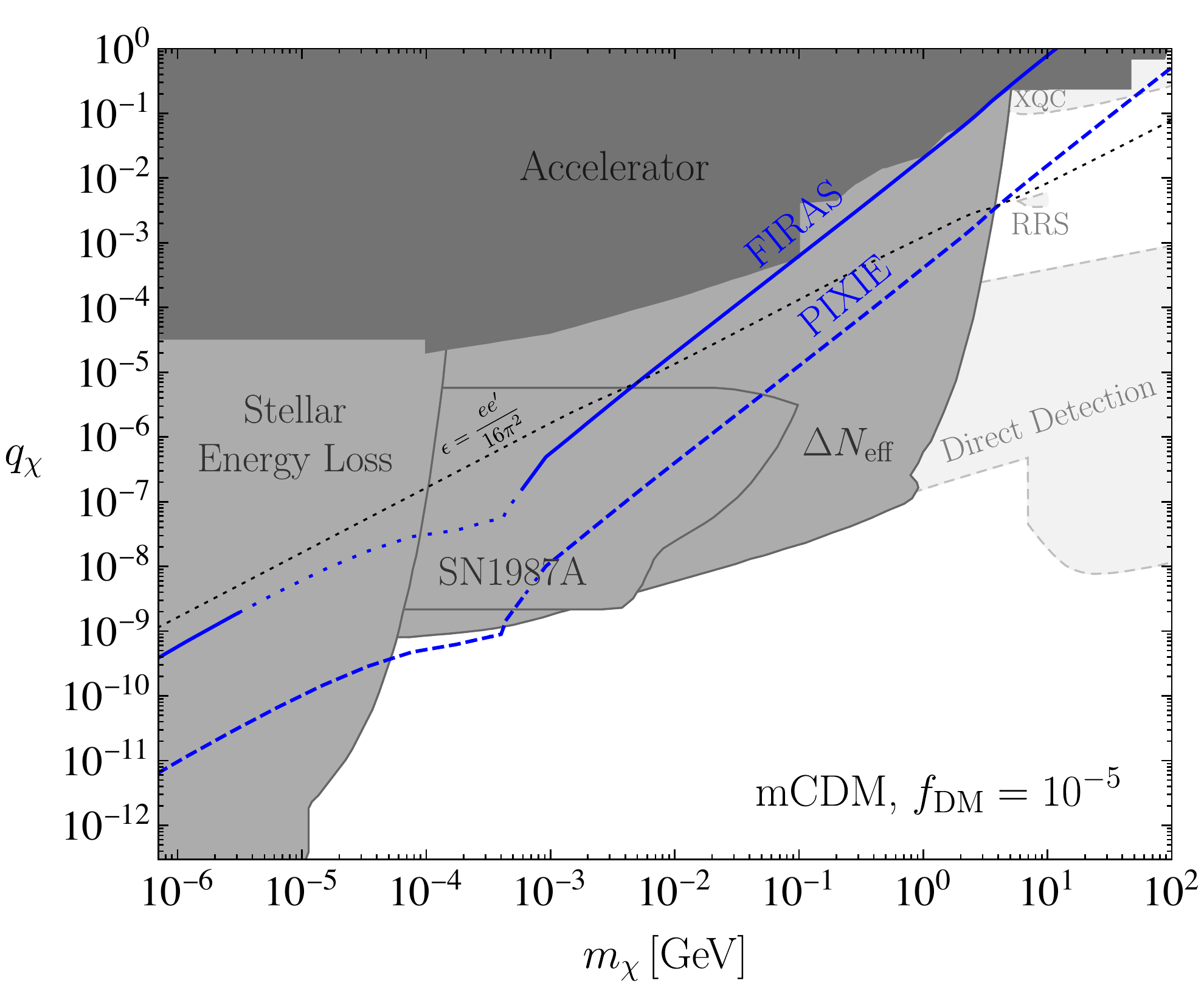}
\vspace{0.3cm}
\caption{Existing limits (solid blue) and future projections (dashed blue) on millicharged dark matter from existing and upcoming measurements of the CMB spectrum by COBE/FIRAS and PIXIE, respectively. Other constraints are shown as shaded gray regions~\cite{Prinz:1998ua, Davidson:2000hf, Magill:2018tbb, ArgoNeuT:2019ckq, Kelly:2018brz, Foroughi-Abari:2020qar, milliQan:2021lne, Kim:2021eix, Kling:2022ykt, Plestid:2020kdm, ArguellesDelgado:2021lek, Vogel:2013raa, Vinyoles:2015khy, Hambye:2018dpi, Chang:2018rso, Mahdawi:2018euy, Emken:2019tni, Adshead:2022ovo}. Limits that rely on the assumption that millicharged dark matter resides in the galactic disk are shown as light dashed gray regions~\cite{Chuzhoy:2008zy,McDermott:2010pa, Hambye:2018dpi,Emken:2019tni}. The parameter space is presented as a function of effective millicharge $q_\x = \eps \, e^\p / e$ and dark matter mass $m_\x$. The dark gauge coupling is fixed as a function of the dark matter mass, assuming a standard thermal freeze-out abundance from $\x^+ \x^- \to \Ap \Ap$, according to Eqs.~(\ref{eq:alphaDFO}) and (\ref{eq:wpFO}). Along the dotted black line, $\eps = e e^\p / 16 \pi^2$ is fixed to the value expected in a minimal model of radiatively-induced kinetic mixing. In each panel, we take the millicharge relics to constitute a fraction $f_\text{DM} = 0.4 \times 10^{-2}$ (top-left panel), $10^{-4}$ (top-right panel), and $10^{-5}$ (bottom panel) of the total dark matter abundance.}
\label{fig:mCDM}
\end{figure}

One of the simplest scenarios of mCDM corresponds to when mCPs arise as a thermal relic of the early universe. This naturally occurs for $q_\x \gtrsim 10^{-7} \times (m_\x  / \GeV)^{1/2}$, such that annihilations of electrons into mCPs equilibrate the SM and dark sectors at early times. In this case, the depletion of the thermal mCP density is typically governed by the freeze-out of annihilations into massless dark photons, $\x^+ \x^- \leftrightarrow \Ap \Ap$, which easily dominates over annihilations into pairs of SM particles for $e^\p / e \gg \eps$ (this is the case for the entire parameter space of interest). These processes deplete the total mCDM density to fractional DM abundances $f_\text{DM}$ for couplings of roughly
\be
\label{eq:alphaDFO}
\alpha^\p \sim 10^{-4} \times \bigg( \frac{ f_\text{DM} }{ 10^{-2}} \bigg) ^{-1/2} \, \bigg( \frac{ m_\x }{  1 \ \GeV } \bigg) 
~.
\ee
To derive \Eq{alphaDFO}, we employ the standard freeze-out estimate $\langle  \sigma v \rangle_{\chi^+ \chi^-} \sim \alpha'^2/m_\chi^2 \sim 1/ (T_\text{eq} \, m_\text{pl})$, where $T_\text{eq} \sim 1 \ \text{eV}$ is the temperature at matter radiation equality and $\mpl$ is the Planck mass. In our analysis, we go beyond this order of magnitude estimate and employ a semi-analytic solution to the Boltzmann equation, as in Ref.~\cite{Chu:2011be,Hambye:2019dwd,Agrawal:2016quu}. As discussed above, this mCP density contributes an in-medium correction to the dark photon's dispersion relation. For values of the dark coupling fixed to the freeze-out estimate in \Eq{alphaDFO}, the $\Ap$ effective mass at redshift $z$ is approximately
\be
\label{eq:wpFO}
\meff^\p \sim 10^{-17} \ \eV \times (1+z)^{3/2} \, \bigg( \frac{ m_\x }{  1 \ \GeV } \bigg) ^{-1/2}  \, \bigg( \frac{ f_\text{DM} }{10^{-2}} \bigg) ^{1/4}
~.
\ee
Using this expression, we can use FIRAS data to place limits on thermal freeze-out cosmologies of mCDM\@. More specifically, for a given mass $m_\x$ and fractional abundance $f_\text{DM}$, we fix $\alpha^\p$ according to the cosmologically-motivated value in \Eq{alphaDFO}, which in turn is used to determine the dark photon's effective mass, as in \Eq{wpFO}. The rest of the procedure is the same as that described in \Sec{cmb} to place an upper limit on the kinetic mixing $\eps$, which can be reexpressed as a limit on the effective charge $q_\x = \eps \, e^\p / e$. The resulting FIRAS limits are shown as solid blue lines in \Fig{mCDM} for the parameter space spanned by $q_\x$ and $m_\x$ and particular choices of $f_\text{DM} = 0.4 \times 10^{-2}$ (top-left panel), $10^{-4}$ (top-right panel), and $10^{-5}$ (bottom panel).

Note that \Eq{wpFO} implies that there are no resonant $\g \leftrightarrow \Ap$ transitions for fixed $f_\text{DM}$ and very light mCPs, since in this case $\meff^\p \gtrsim \w_p$ at all redshifts for sufficiently small masses. This corresponds to the sharp edge at the left-most part of the sensitivity regions in \Fig{mCDM}, which scales with the DM fraction as $\propto f_\text{DM}^{1/2}$ (this is visually apparent within the mass range shown only for the top-left panel of \Fig{mCDM}). 
In our calculations, we have checked that the mCP self-interaction rate satisfies $\Gamma^\p \ll \w$ across the entire parameter space shown, allowing us to approximate $ \meff^\p \simeq \w_p^\p $. Projections from the proposed PIXIE satellite are also shown in dashed blue (see \Sec{cmb} for additional details), which can improve upon the reach compared to FIRAS by roughly two orders of magnitude in the effective charge $q_\x$. Also shown in \Fig{mCDM} as a dotted black line is the value of the kinetic mixing parameter that is motivated by simple models of radiatively-induced coupling, corresponding to $\eps = e e^\p / 16 \pi^2$ or equivalently $q_\x = \alpha^\p / 4 \pi$~\cite{Holdom:1985ag}. Various theoretical constructions can populate models above or below this line. For instance, dark sectors involving an enhanced dark charge/flavor number correspond to $q_\x \gg \alpha^\p / 4 \pi$, whereas non-abelian dark sectors or those generating multi-loop kinetic mixing correspond to $q_\x \ll \alpha^\p / 4 \pi$~\cite{Arkani-Hamed:2008kxc,Gherghetta:2019coi}.

Previously investigated limits, such as those derived from cosmology, astrophysics, and terrestrial searches are shown in gray. The darkest gray regions are excluded from accelerator and neutrino experiment searches~\cite{Prinz:1998ua,Magill:2018tbb,ArgoNeuT:2019ckq,Kelly:2018brz,Foroughi-Abari:2020qar,milliQan:2021lne, Kim:2021eix, Kling:2022ykt, Davidson:2000hf,ArguellesDelgado:2021lek}. The next lightest gray shaded regions are excluded from considerations of stellar energy loss~\cite{Davidson:2000hf,Vogel:2013raa,Vinyoles:2015khy}, SN 1987A~\cite{Chang:2018rso}, and early universe probes of the modification to the effective number of neutrino species $\Delta N_\text{eff}$  (resulting from the massless dark photon's contribution to the radiation energy density)~\cite{Vogel:2013raa,Adshead:2022ovo}. We note that for certain regions of parameter space mCPs remain thermalized with the SM at sufficiently low temperatures such that this energy density in dark photons is comparable to that of the SM photon density. As mentioned above in \Sec{transitions}, in this case the inverse process $\Ap \to \g$ can become comparable or dominate over $\g \to \Ap$, modifying the form of the spectral distortions discussed so far. We refrain from performing a careful estimate of such effects since this part of parameter space is typically in tension with cosmological probes of $\Delta N_\text{eff}$.  This can be seen as the dotted part of the otherwise solid ``FIRAS" line in \Fig{mCDM}, which highlights where our calculated limits overlap this region of parameter space and thus are expected to be modified. 

In the light dashed gray regions, we show limits from searches for energy depositions from mCDM scattering on terrestrial detectors, such as underground and surface-level DM direct detection experiments~\cite{Emken:2019tni, Hambye:2018dpi} and the balloon- and rocket-based calorimeter experiments RRS and XQC~\cite{Mahdawi:2018euy}. It is important to note that such signals require mCDM to reside locally in the Milky Way, whereas signals of spectral distortions rely on the cosmological presence of mCPs at early times. Along these lines, previous studies have suggested that in this parameter space supernovae remnants may have evacuated mCDM from the galactic disk, rendering terrestrial DM searches insensitive to mCP relics~\cite{Chuzhoy:2008zy,McDermott:2010pa,Sanchez-Salcedo:2010gfa}.  In this case, FIRAS is the strongest probe of the simplest incarnations of mCDM for $m_\x \gtrsim 1 \ \GeV$ and $10^{-5} \lesssim f_\text{DM} \lesssim 10^{-2}$. 

The above discussion demonstrates that measurements of the CMB spectrum are sensitive to new sets of models involving massless dark photons and thermal mCP relics. The latter assumption was critical in fixing the value of $\alpha^\p$ and allowing us to restrict to the three-dimensional parameter space spanned by $q_\x$, $m_\x$, and $f_\text{DM}$. One could alternatively consider other thermal histories that would produce a sizable mCP number density with larger $ \alpha^\p$ (e.g., if there exists a  particle-antiparticle asymmetry in the mCP population). This would further enhance the importance of CMB spectrum measurements relative to the other constraints shown in gray in \Fig{mCDM}.

Before proceeding, we note that Ref.~\cite{Bogorad:2021uew} recently investigated similar constraints on bosonic mCDM\@. That work focused on much lighter ($\ll 1 \ \MeV$) and much more weakly coupled ($q_\x \ll 10^{-10}$) mCPs produced from the misalignment mechanism before the end of inflation and did not consider the potentially important effects arising from self-interactions. Although self-interactions are irrelevant in our work, in the ultralight bosonic regime of Ref.~\cite{Bogorad:2021uew} the increased number density and phase space occupancy may lead to enhanced self-scattering that can suppress the effective dark photon mass, as in \Eq{omega1body}. A detailed investigation of such effects in the ultralight parameter space is beyond the scope of this work.

\subsection{Millicharged Dark Radiation (mCDR)}
\label{sec:DRlimit}

In this section, we investigate constraints on relativistic mCPs, which contribute to the cosmological radiation energy density. In this limit, the mCP mass is assumed to be sufficiently small to be irrelevant, such that from \Eq{wpGen} the dark plasma frequency is approximately $\w_p^{\p \, 2} \simeq (8 \pi / 3) \alpha^\p n_\x \langle 1/p_\x \rangle$, where the momentum $p_\x$ in the brackets is averaged over the mCP phase space~\cite{Raffelt:1996wa}. From \Eq{omega1body}, this gives rise to an effective $\Ap$ mass-squared of~\footnote{Here, we neglect mCDR self-interactions and set $\Gamma^\p = 0$ since relativistic scattering is not enhanced by a small relative velocity.}
\begin{equation} 
\label{eq:mApDR}
\meff^{\p \, 2} \simeq 4 \pi \, \alpha^\p \, n_\x \langle 1/p_\x \rangle\,.
\end{equation} 
For instance, the phase space of an mCDR sector made up of a particle-antiparticle pair of a complex scalar or a Dirac fermion described by a temperature $T_\x \gg m_\x$ leads to $\meff^{\p \, 2} \simeq (2 \pi / 3) \, \alpha^\p \, T_\x^2$. It is often convenient to parametrize $T_\x$ in terms of the effective additional neutrino degrees of freedom $\Delta \Neff$ in mCDR,
\begin{equation} 
T_\x = \bigg( \frac{4}{11} \bigg)^{1/3} \,  \bigg(  \frac{7}{4} \bigg)^{1/4} \bigg(\frac{\Delta \Neff}{g_*^\x} \bigg)^{1/4} ~ T_\g
~,
\end{equation} 
where $g_*^\x$ is the effective relativistic degrees of freedom of the mCDR and $T_\g$ is the temperature of the SM photon bath~\cite{Kolb:1990vq}. Planck observations of the CMB, in combination with other cosmological and astrophysical datasets, restrict the level of additional free-streaming dark radiation to be $\Delta \Neff \lesssim 0.3$~\cite{Planck:2018vyg}. For concreteness, we focus our attention on fermionic mCDR for the remainder of this section (similar expressions hold for scalar mCDR). In this case, the effective dark photon mass evolves as a function of redshift as  
\begin{equation} 
\label{eq:WpR}
\meff^\p \sim 10^{-12} \ \eV \times (1+z) ~ \bigg( \frac{ \Delta \Neff }{  0.3} \bigg) ^{1/4} ~ \bigg( \frac{ q_\x }{  10^{-14}} \bigg)  ~ \bigg( \frac{ 10^{-7} }{  \eps } \bigg) \qquad (\text{Fermi-Dirac})
~.
\end{equation} 
To derive this expression, we used that $e^\p = e q_\x / \eps$ and normalized $q_\x$ to be slightly smaller than the strongest existing upper bounds on massless mCPs  ($q_\chi \lesssim 2 \times 10^{-14}$), arising from considerations of stellar energy loss~\cite{Davidson:2000hf,Vogel:2013raa,Vinyoles:2015khy}.
\begin{figure}[t]
\centering
\begin{tikzpicture} 
\node at (-7.5,0){\includegraphics[width=0.50\columnwidth]{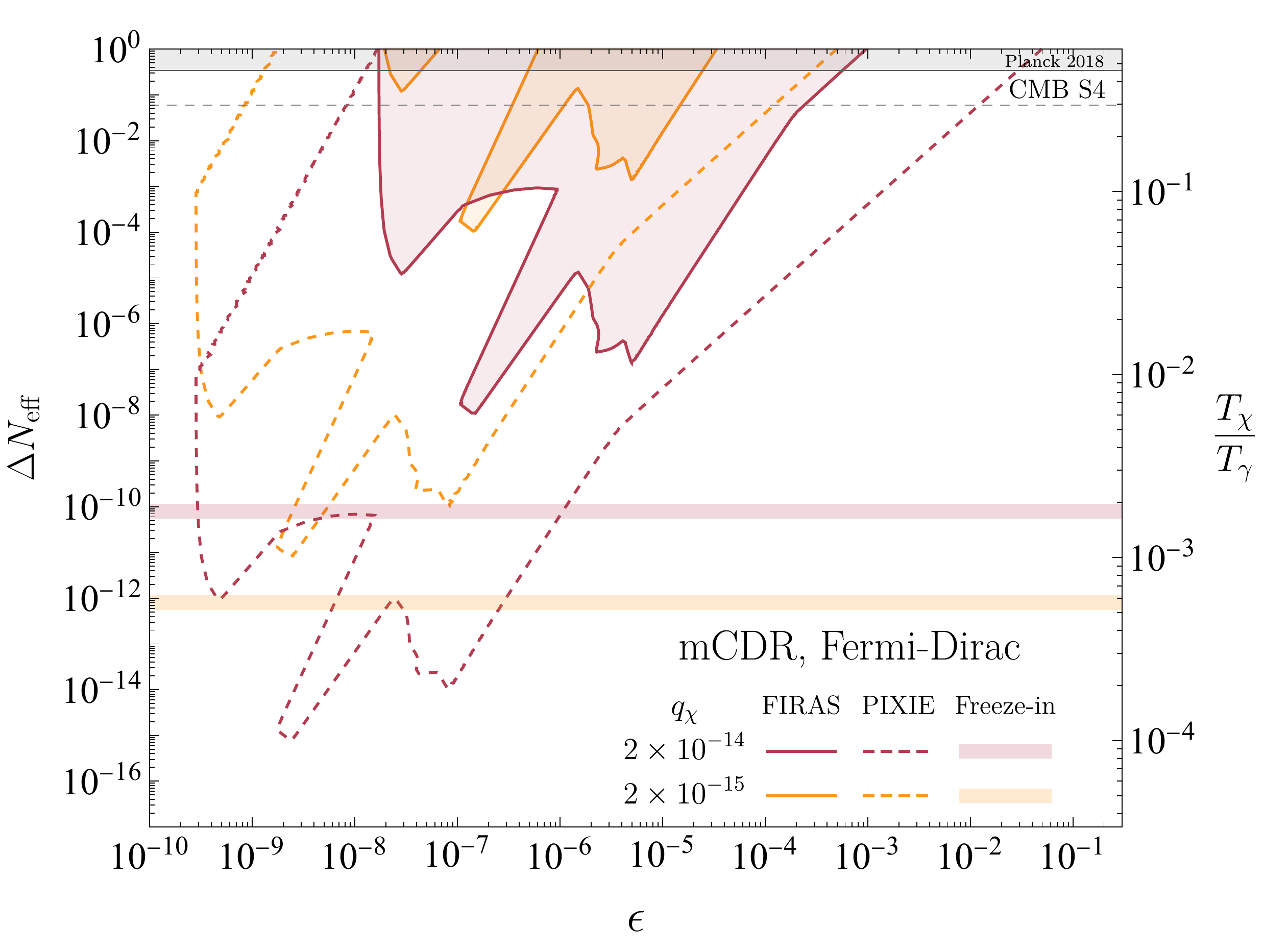}};
\node at (1.5,0.0) {\includegraphics[width=0.457\columnwidth]{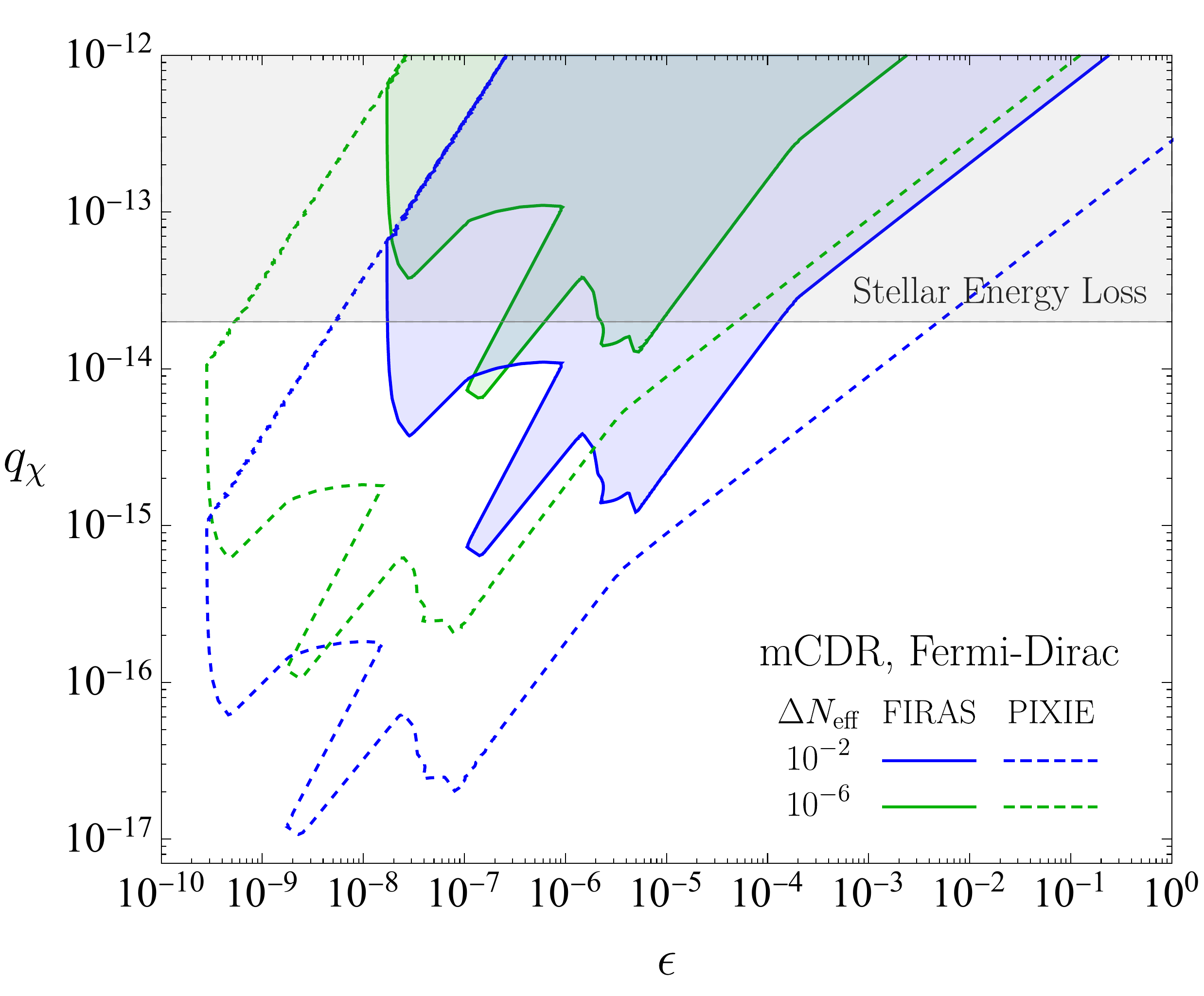}};
\end{tikzpicture}
\vspace{0.3cm}
\caption{Existing limits (solid) and future projections (dashed) on millicharged dark radiation from existing and upcoming measurements of the CMB spectrum by COBE/FIRAS and PIXIE, respectively. Here, we assume that the millicharge phase space is well-approximated by a thermal Fermi-Dirac distribution. {\bf Left}: The parameter space is presented as a function of effective additional neutrino degrees of freedom in millicharged particles $\Delta \Neff$ and kinetic mixing $\eps$ for two representative values of the effective charge $q_\x = 2 \times 10^{-14}$ (red) and $2 \times 10^{-15}$ (orange). Note that $\Delta \Neff$ can be translated into a dark sector-to-Standard Model temperature ratio $T_\x / T_\g$, as shown on the right axis. For each choice of $q_\x$, we show a horizontal band corresponding to the amount of millicharged dark radiation predicted to be produced by freeze-in via plasmon decay, $\g^* \to \x^+ \x^-$. Also shown as a shaded gray region or dotted gray line are existing constraints or future projections on additional dark radiation from CMB observations by Planck~\cite{Planck:2018vyg} and CMB-S4~\cite{Abazajian:2019eic}, respectively.  {\bf Right}: The parameter space is presented as a function of the effective millicharge $q_\x = \eps \, e^\p / e$ and kinetic mixing $\eps$ fixing $\Delta \Neff = 10^{-2}$ (blue) and $10^{-6}$ (green). The shaded gray region is excluded from considerations of stellar energy loss~\cite{Davidson:2000hf,Vogel:2013raa,Vinyoles:2015khy}.}
\label{fig:mCDR}
\end{figure}

The discussion above details the mapping of the mCDR model parameters to the effective $\Ap$ mass. In \Fig{mCDR}, we use this mapping to recast the mCDR bounds from \Fig{zevolution} to the mCP parameter space. For an mCDR population well-described by a thermal Fermi-Dirac distribution, the model is completely specified by $\Delta \Neff$  (or equivalently the temperature ratio $T_\x / T_\g$) and the couplings $\eps$ and $q_\x$. In the left panel of \Fig{mCDR}, we fix the effective charge $q_\x$ to be either the largest value allowed by existing constraints on ultralight mCPs (red)~\cite{Davidson:2000hf,Vogel:2013raa,Vinyoles:2015khy}, or an order of magnitude smaller (orange), and show the existing limits or future projections from COBE/FIRAS (solid lines) and PIXIE (dashed lines), respectively, as a function of $\Delta \Neff$ and $\eps$. Note that on the right vertical axis, for each choice of $\Delta \Neff$ we show the corresponding value of the mCP-to-photon temperature ratio $T_\x / T_\g$. Also shown as a shaded gray region or dotted gray line are existing constraints or future projections on additional dark radiation from CMB observations by Planck and CMB-S4, respectively~\cite{Planck:2018vyg,Abazajian:2019eic}. Instead, in the right panel we fix $\Delta \Neff$ to two representative values of $10^{-2}$ (blue) or $10^{-6}$ (green) and display the parameter space spanned by the couplings $q_\x$ and $\eps$. The shaded gray region is excluded from considerations of stellar energy loss~\cite{Davidson:2000hf,Vogel:2013raa,Vinyoles:2015khy}. 

\Fig{mCDR} illustrates that spectral measurements of the CMB are sensitive to even tiny amounts of additional radiation consisting of extremely feebly-coupled mCPs. In light of these findings, it is worth investigating the simplest cosmologies that give rise to such small amounts of additional dark radiation, analogous to the philosophy outlined above in \Sec{DMlimit}\@. Infrared ``freeze-in" is an especially compelling example of such a cosmology, as it is insensitive to much of the dynamics in the early universe and provides a semi-predictive mapping between $q_\x$ and $\Delta \Neff$~\cite{Hall:2009bx}. In particular, the abundance of ultralight mCPs has an irreducible contribution from the decay of SM plasmons $\g^* \to \x^+ \x^-$ with a rate controlled by the size of $q_\x$~\cite{Dvorkin:2019zdi}. As discussed in \App{freezein}, we find that the comoving number of mCDR produced from such freeze-in processes is approximately
\begin{equation} 
\label{eq:freezeinyield}
Y_\x^{\text{(FI)}} = \frac{ n_\x^{\text{(FI)}}}{  s_\g } \simeq 3 \times 10^{-12} \times \bigg( \frac{ q_\x }{  10^{-14}}\bigg)^2
~,
\end{equation} 
where $s_\g \simeq 1.7 \times T_\g^3$ is the entropy density of the visible sector at a temperature $T_\g$ below the electron mass threshold. The corresponding energy density in this mCDR is approximately
 \begin{equation}  
\label{eq:freezeinNeff}
\Delta \Neff^{\text{(FI)}} \simeq 2 \times 10^{-11}  \times \bigg( \frac{ q_\x }{  10^{-14}} \bigg)^2
~.
\end{equation} 
The horizontal bands in the left panel of \Fig{mCDR} show this predicted value of $\Delta \Neff$ from freeze-in for $q_\x = 2 \times 10^{-14}$ (red) or $q_\x = 2 \times 10^{-15}$ (orange), assuming that freeze-in from plasmon decay is the only process contributing to the cosmological abundance of mCDR\@. Hence, for these choices of $q_\x$, regions above the corresponding horizontal lines require additional dynamics in the early universe to produce mCPs, whereas regions below require new processes that deplete the mCP abundance below that generated by plasmon decay. 

\Fig{mCDR} assumes that the mCP phase space is well-described by a thermal Fermi-Dirac distribution. However, this is not guaranteed to be the case. For instance, the resulting freeze-in contribution to the effective dark photon mass depends crucially on the post-production evolution of the mCP phase space. This is due to the fact that the typical energy $p_\x \sim T_\g$ of an ultra-relativistic mCP produced from plasmon decay is large compared to what would be expected from its number density, i.e., $\langle p_\x \rangle^3 / n_\x^{\text{(FI)}} \sim 1/Y_\x^{\text{(FI)}} \gg 1$; in other words, the phase space is highly non-thermal~\cite{Dvorkin:2019zdi}. As a result, this high-momentum phase-space tail suppresses the contribution to the effective $\Ap$ mass in \Eq{mApDR}, compared to a thermal distribution. Assuming that this phase space remains unaltered after freeze-in, we find that the in-medium contribution to the dark photon mass is
\begin{equation} 
\label{eq:OmegaFreezein}
\meff^\p  \simeq 3 \times 10^{-17} \ \eV \times (1+z) ~ \bigg( \frac{ q_\x }{  10^{-14}} \bigg)^2 ~ \bigg( \frac{ 10^{-7}  }{  \eps }\bigg) \qquad (\text{freeze-in})
~.
\end{equation} 
In \App{freezein}, we derive \Eq{OmegaFreezein} by directly solving the Boltzmann equation.
\begin{figure}[t]
\centering
\includegraphics[width=0.6\columnwidth]{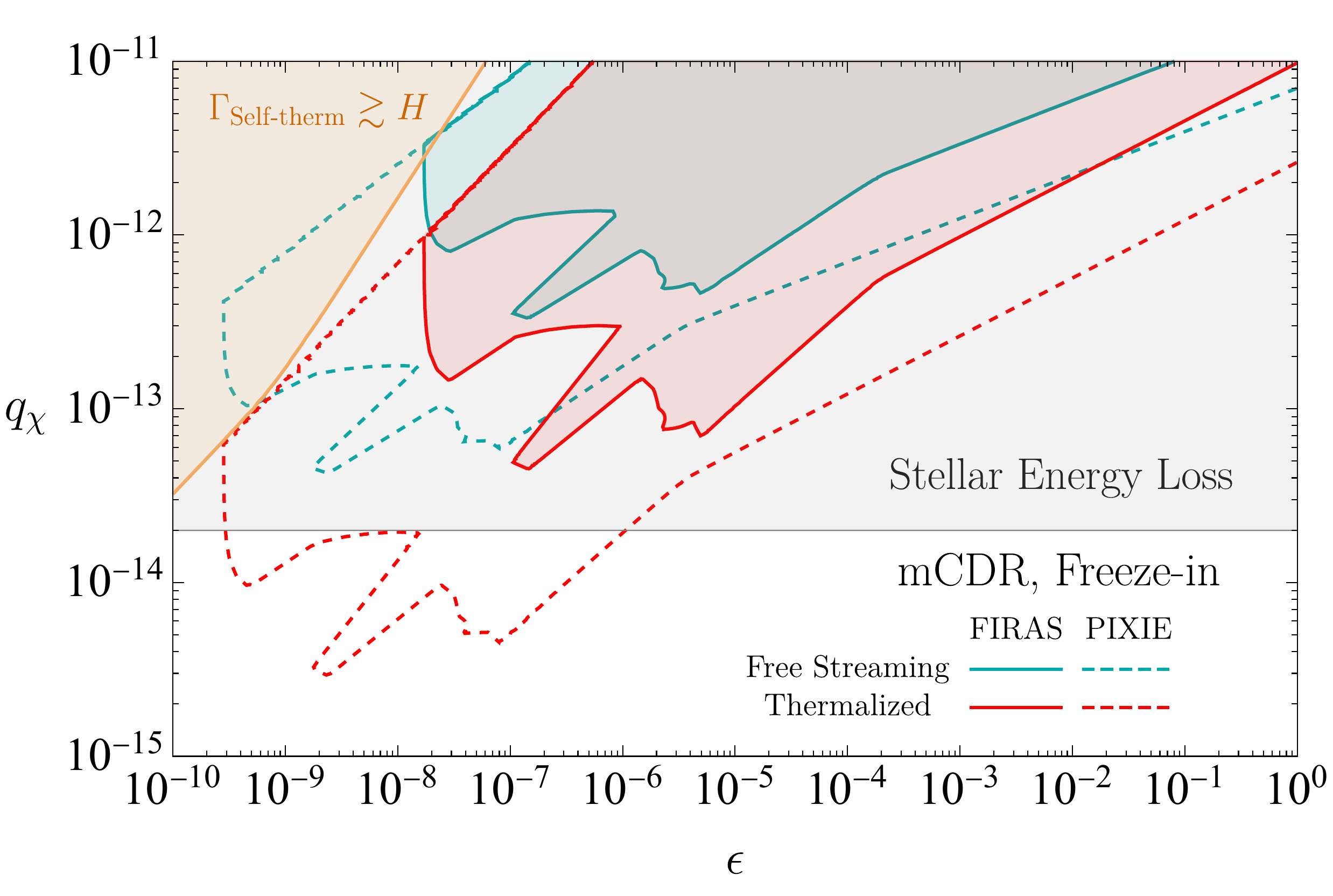}
\vspace{0.3cm}
\caption{Limits (solid) and projections (dashed) on millicharged dark radiation that is produced through freeze-in via plasmon decay $\g^* \to \x^+ \x^-$, as in Eqs.~(\ref{eq:freezeinyield})$-$(\ref{eq:OmegaFreezein}), from existing and upcoming measurements of the CMB spectrum by COBE/FIRAS and PIXIE, respectively. The results are shown assuming that either the millicharged particle phase space distribution is unperturbed compared to that generated from freeze-in production (cyan) or that self-interactions drive the phase space towards that of a thermal Fermi-Dirac distribution (red). In the orange shaded region, $\x^+ \x^- \to \Ap \Ap$ reactions modify the dark sector phase space before the last resonant $\g \to \Ap$ transition; outside this region self-thermalization requires the presence of additional unspecified interactions within the dark sector. The shaded gray region is excluded from considerations of stellar energy loss~\cite{Davidson:2000hf,Vogel:2013raa,Vinyoles:2015khy}.}
\label{fig:freezein}
\end{figure}

The late-time phase space is significantly modified if dark sector number-changing and scattering processes become efficient, such as $\x^+ \x^- \to \Ap \Ap$, $\x \Ap \to \x \Ap$, and $\chi A' \rightarrow \chi A' A'$. Since $\x^+ \x^- \to \Ap \Ap$ is the dominant process, we conservatively estimate that the freeze-in phase space is significantly modified provided that $\alpha^{\p \, 2} \, n_\x^{\text{(FI)}} / T_\g^2 \gg H$, where $n_\x^{\text{(FI)}}$ is the number density of mCPs generated from plasmon decay. From this estimate, we find that the distribution of an mCDR population initially populated from freeze-in is significantly altered before a temperature $T_\g$ for couplings of
\begin{equation} 
\label{eq:DRtherm}
q_\x  \gg 10^{-11} \times \bigg( \frac{ \eps }{  10^{-7} } \bigg)^{2/3} ~ \bigg( \frac{ T_\g  }{  \keV} \bigg)^{1/6}
~.
\end{equation} 
When \Eq{DRtherm} is satisfied, the mCDR population begins to approach that of a thermal Fermi-Dirac distribution. We leave a more complete calculation to determine the precise behavior of the mCP phase space to future work. If self-thermalization occurs, the temperature $T_\x$ of this newly equilibrated population is determined by energy conservation, i.e., $n_\x^{\text{(FI)}} \, T_\g \sim T_\x^4$. We therefore estimate that the resulting in-medium contribution to the $\Ap$ mass is enhanced compared to that of \Eq{OmegaFreezein} by
\begin{equation} 
\meff^\p (\text{self-thermalized}) \sim\meff^\p (\text{freeze-in}) \times \big( Y_\x^{\text{(FI)}} \big)^{-1/4}
~,
\end{equation} 
where $Y_\x^{\text{(FI)}} \ll 1$ is the yield determined by \Eq{freezeinyield}. As expected, self-thermalization of the dark sector parametrically increases the dark photon's effective mass. 

\Fig{freezein} investigates the minimal cosmological scenario where freeze-in from plasmon decay is solely responsible for the production of mCPs. Hence, the initial density is fixed as a function of $q_\x$ by \Eq{freezeinyield}. We then calculate the resulting constraints and projections from COBE (solid lines) and PIXIE (dashed lines) for two scenarios: where this mCP population either self-thermalizes (red) or retains its original distribution (cyan). In the region labelled $``\Gamma_\text{Self-therm} \gtrsim H$," $\x^+ \x^- \to \Ap \Ap$ modifies the mCP phase space before the latest resonantly enhanced $\g \to \Ap$ transition occurs. Below this region, additional forms of self-interactions are required to self-thermalize the dark sector. From \Fig{freezein} we note that the irreducible contribution to the mCP density from plasmon decay is typically too small to give rise to sufficiently large spectral distortions as to be competitive with existing stellar constraints. Regardless, PIXIE would be sensitive to new freeze-in parameter space in the case that the mCDR self-thermalizes through some unspecified additional self-interaction.

\section{Opening Up the Parameter Space for Massive Dark Photons }
\label{sec:darkphoton}
Above, we investigated the prospects of detecting unbroken symmetries using the presence of mCPs. In doing so, we placed new limits within the mCP parameter space, assuming the simplest cosmologies for  mCDM and mCDR\@. Alternatively, one may ask how the usual bounds on massive dark photons as derived from $\g \rightarrow \Ap$ oscillations are altered in the presence of mCP relics. More specifically, we seek to evaluate the robustness of such bounds in regards to minor modifications to the cosmic history. In the presence of mCPs, massive dark photons have two contributions to their effective mass, which can be incorporated by replacing $\meff^{\p \, 2} \to \mAp^2 + \meff^{\p \, 2}$ in the results presented above, where $\mAp$ is the usual in-vacuum dark photon mass and $\meff^\p$ is the in-medium mCP contribution. The redshift dependence of $\meff^\p$ depends on whether or not the mCP relics are relativistic (see, e.g., \Eq{meffgen}). As we discuss below, the steep scaling $\meff^\p \propto (1+z)^{3/2}$ for non-relativistic mCPs implies that mCDM can significantly weaken existing cosmological constraints on massive dark photons. On the other hand, the weaker scaling $\meff^\p \propto (1+z)$ for relativistic mCPs implies that mCDR typically only strengthens such bounds. Since this latter point was already investigated in \Sec{DRlimit}, we focus solely on mCDM in this section. 

\begin{figure}[t]
\centering
\begin{tikzpicture} 
\node at(-7.5,0){\includegraphics[width=0.46\columnwidth]{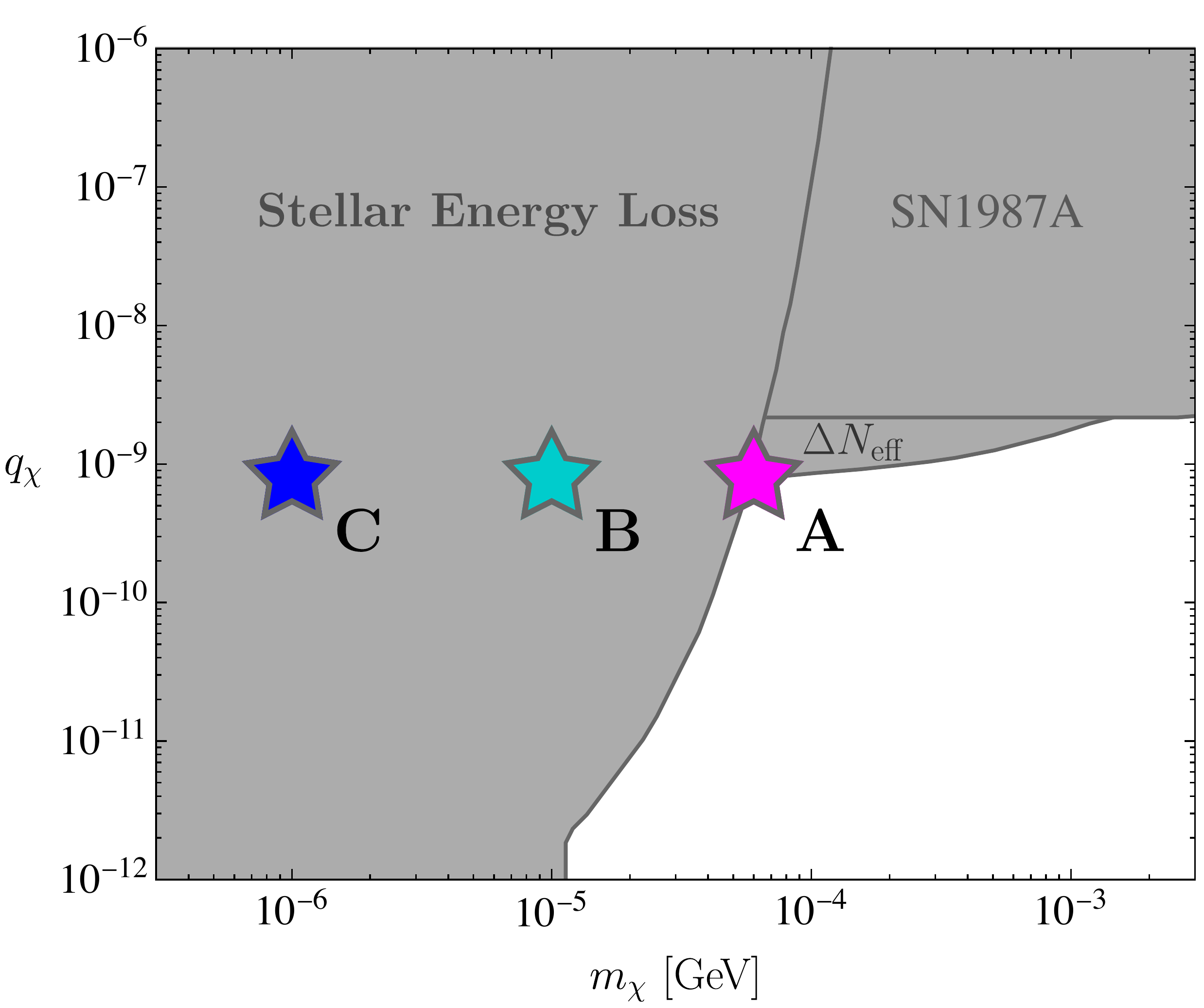}};
\node at (1.25,0){\includegraphics[width=0.4475\columnwidth]{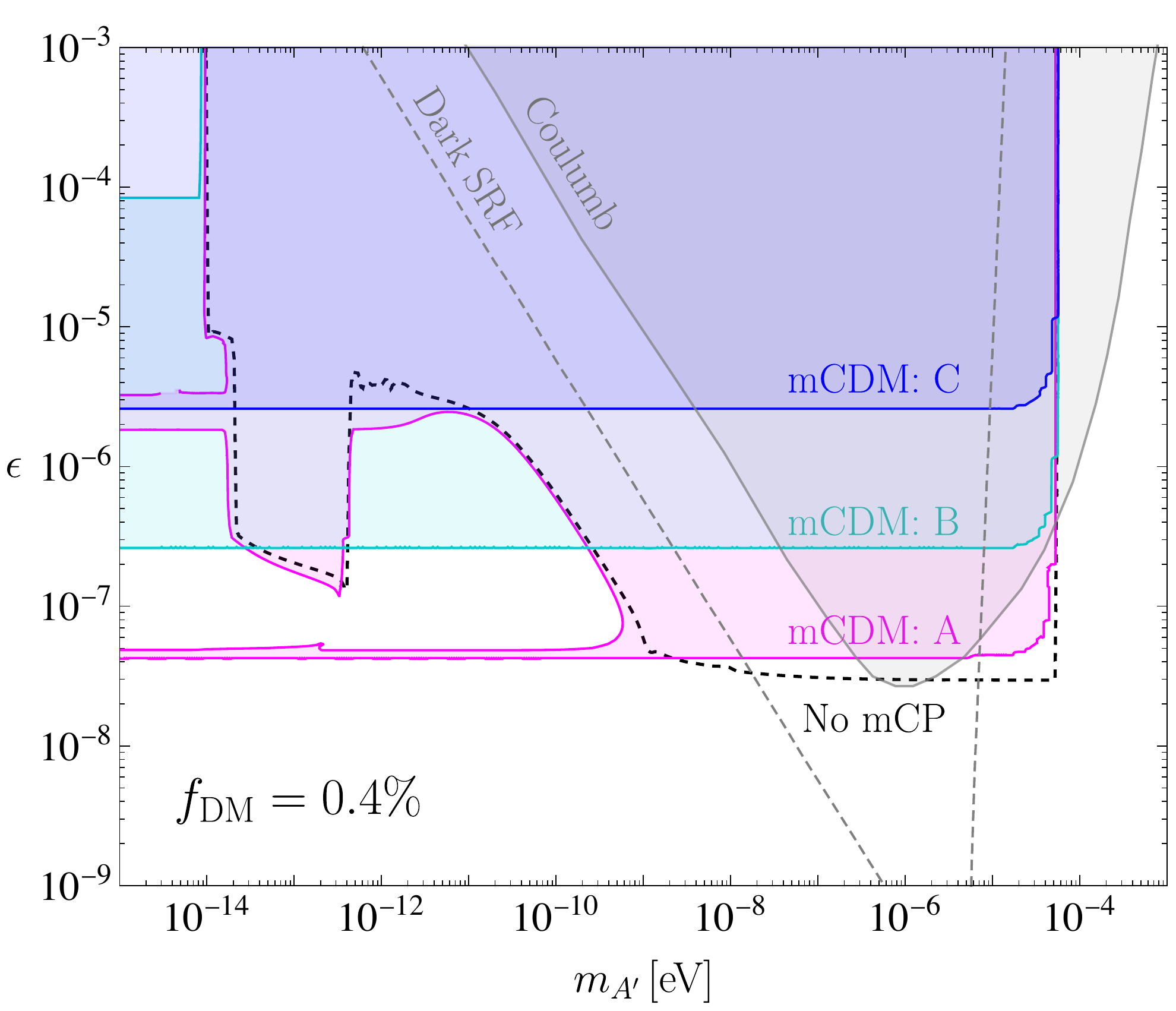}};
\end{tikzpicture}
\vspace{0.3cm}
\caption{The modification of bounds on massive dark photons in the presence of millicharged particle dark matter subcomponents, fixing the fractional abundance to $f_\text{DM} = 0.4\%$. {\bf Left}: The three chosen  benchmark points in the millicharged particle parameter space, spanned by the charge $q_\x$ and mass $m_\x$. These three benchmark points are chosen in light of known outstanding questions regarding the robustness of stellar energy loss constraints~\cite{Davidson:2000hf,Vogel:2013raa,Vinyoles:2015khy}. Point A (magenta star) is not excluded by any existing constraints. Point B (cyan star) lies within the mass range where calculations of stellar energy loss require a precise modeling of the stellar temperature profile. Point C (blue star) corresponds to much smaller masses such that millicharged particle production is not exponentially sensitive to the stellar temperature; however, for such large couplings, the gyroradius of a millicharged particle within the stellar magnetic field is much smaller than the stellar radius, thus requiring a dedicated analysis beyond present estimates~\cite{Vinyoles:2015khy,AndreaFuture}. {\bf Right}: Bounds from COBE/FIRAS on massive dark photons for each of the model points shown in the left panel (points A, B, and C correspond to the shaded magenta, cyan, and blue regions, respectively). For reference, we also show the usual bound assuming a negligible density of millicharged relic particles (dashed black). In shaded and dotted gray, we also show existing constraints from previous tests of Coulomb's law~\cite{Bartlett:1970js, Williams:1971ms,Bartlett:1988yy} and a future projection of the DarkSRF light-shining-through-wall experiment at Fermilab~\cite{Berlin:2022hfx}, respectively.}
\label{fig:Aprime_NR}
\end{figure}

The mCDM contribution to the dark photon effective mass $\meff^\p$ has the same redshift scaling as the free electron contribution to the SM photon plasma mass $\w_p$. Furthermore, since the effect of neutral hydrogen is only to reduce $\w_p$, if $\meff^\p$ is sufficiently large it will entirely prevent resonant $\gamma \rightarrow \Ap$ transitions from being imprinted on the CMB\@. This effect is most pronounced for very light mCDM, since $\meff^\p$ scales inversely with the DM mass for fixed mCP fractional density $f_\text{DM}$. We hence choose to focus on the light (sub-MeV) mCDM parameter space, since this maximizes the possible modifications to the dark photon's effective mass.  The relevant mCP parameter space is displayed in the left panel of \Fig{Aprime_NR}. As opposed to the $m_\x \gtrsim \MeV$ parameter space investigated in Sec.~\ref{sec:DMlimit}, cosmological and astrophysical bounds on sub-MeV mCPs requires much smaller couplings, $q_\x \lesssim 10^{-9}$~\cite{Adshead:2022ovo,Chang:2018rso}. Also unlike in Sec.~\ref{sec:DMlimit}, here we refrain from specifying a particular cosmological production mechanism for the mCDM. In fact, for the choices of $e^\p = e q_\x / \eps $ that we consider below, a symmetric mCDM population in the early universe would have either annihilated away via $\x^+ \x^- \to \Ap \Ap$ to negligible densities for $\eps \lesssim 10^{-6}$, or would have been overproduced by $\g^* \to \x^+ \x^-$ for $\eps \gtrsim 10^{-6}$. Hence, we implicitly assume the presence of additional dynamics corresponding to, e.g., an initial cosmological mCP asymmetry or an additional mCP annihilation channel, depending on the particular value of $\eps$.

Fixing $m_\x = 50 \ \keV$, $q_\x = 10^{-9}$, and $f_\text{DM} = 10^{-2}$ in \Eq{wp}, we have that the mCDM contribution to the effective $\Ap$ mass at redshift $z = 0$ is $\meff^\p (0) \sim \text{few} \times 10^{-14} \ \eV \times  (10^{-7}/\eps)$ which is slightly greater than $\w_p$ at $z = 0$ and is hence large enough to prevent $\gamma \rightarrow \Ap$ resonant transitions at all redshifts. In this section, we study the influence of mCDM on the parameter space of massive dark photons fixing $q_\x = 8 \times 10 ^{-10}$ (the largest cosmologically-allowed coupling in this low mass parameter space~\cite{Adshead:2022ovo}), $f_\text{DM} = 0.4\%$ (the largest allowed fraction for tightly-coupled mCDM~\cite{dePutter:2018xte,Kovetz:2018zan}), and the mCP mass to one of three values:  $m_\x = 60 \ \keV$ (scenario A), $m_\x = 10 \ \keV$ (scenario B), or $m_\x =  1 \ \keV$ (scenario C). These points are shown as the pink, cyan, and blue stars in the left panel of \Fig{Aprime_NR}, respectively. As seen from \Fig{Aprime_NR}, Point A is allowed by all existing bounds and represents the most conservative choice of model parameters. Points B and C lie within regions of parameter space that are naively excluded by considerations of stellar energy loss~\cite{Davidson:2000hf,Vogel:2013raa,Vinyoles:2015khy}. However, there exist two reasons to question the robustness of such bounds on models of mCPs since there are outstanding questions regarding their reliability in light of stellar modeling of temperature profiles and magnetic fields. First, for masses $m_\x \gtrsim 10 \ \keV$, $m_\x$ is greater than the characteristic temperature of stellar cores, such that emission of  mCPs is exponentially sensitive to the modeling of the stellar interior. This motivates the investigation of parameter point B\@. Second, for couplings $q_\x \gtrsim 10^{-11}$, the gyroradius of an mCP in the presence of stellar magnetic fields is typically much smaller than the size of the star~\cite{Vinyoles:2015khy}, in which case mCPs do not freely propagate outwards, possibly resulting in modifications to the naive bound~\cite{AndreaFuture}. This motivates the investigation of parameter point C.

The COBE/FIRAS limits on massive dark photons from $\g \to \Ap$ oscillations are shown in the right panel of  Fig.~\ref{fig:Aprime_NR} as a function of the dark photon in-vacuum mass $\mAp$ and the kinetic mixing $\eps$ for our three chosen mCDM scenarios. For comparison, in dotted black we also show the standard bound assuming no cosmological abundance of mCPs~\cite{Mirizzi:2009iz,Mirizzi:2009nq,Kunze:2015noa}. Since the mCP contribution to the effective $\Ap$ mass scales as $\meff^\p \propto e^\p \propto q_\x / \eps$, for fixed millicharge $q_\x$ mCPs prevent resonant $\g \to \Ap$ transitions entirely for sufficiently small $\eps$ (corresponding to sufficiently large $e^\p$). On the other hand, for fixed $q_\x$ resonant transitions are induced for sufficiently large $\eps$ with a probability that depends on the size of $\eps$ and $\meff^\p$ (see \Eq{rate3}). Relatedly, in addition to weakening the bounds on massive dark photons at small $\eps$, the presence of mCPs strengthens limits at small dark photons masses, which are otherwise unbounded. These effects are most pronounced for parameter point C, corresponding to the lowest mCP mass considered. We note that we do not consider regions of parameter space corresponding to couplings below upper bounds derived from stellar energy loss, $q_\x \lesssim 10^{-14}$~\cite{Davidson:2000hf,Vogel:2013raa,Vinyoles:2015khy}, although in principle a smaller $q_\x$ could be compensated for by a smaller mCP mass, e.g., $m_\x \ll 1 \ \eV$\@. In this case additional complications potentially arise from large self-interactions (thereby suppressing the mCP contribution to $\meff^\p$) and cosmological instability from parametric resonance~\cite{Jaeckel:2021xyo} (which suppresses the overall mCDM density).

Interestingly, the existence of light mCPs can significantly weaken cosmological bounds on massive dark photons at small values of $\eps$. In such regions of parameter space, terrestrial searches for dark photons thereby play an increasingly important role. In the right panel of Fig.~\ref{fig:Aprime_NR}, the shaded gray region corresponds to a limit derived from a terrestrial test of Coulomb's law~\cite{Bartlett:1970js, Williams:1971ms,Bartlett:1988yy}. Also shown as dotted gray is the projected sensitivity of the light-shining-through-wall experiment DarkSRF at Fermilab~\cite{Berlin:2022hfx}. In this sense, Fig.~\ref{fig:Aprime_NR} highlights the importance to investigate parameter space that is cosmologically excluded in only the simplest model constructions; slight variations to the model can open up previously bounded parameter space, motivating direct terrestrial probes that are largely insensitive to the same model variations.

Before concluding this section, we note that it is worthwhile to consider additional dynamics in the dark sector, beyond that of freely ionized mCPs, since slight modifications to the models investigated above might allow for larger changes to the existing limits on massive dark photons. Along these lines, we have also investigated scenarios where oppositely-charged mCPs bind into neutral states either due to $\Ap$-exchange or through some additional unspecified interactions within the dark sector~\cite{Cyr-Racine:2012tfp,Cline:2012is, Bai:2021nai, Mathur:2021gej}. Analogous to the second term in the first equation of \Eq{omega1body}, dark bound states can be incorporated into our analysis by making the following modification to the effective dark photon mass, $\meff^{\p \, 2} \to   \w_{p, \HI}^{\p \, 2} / (1 - E _0^{\p \, 2} /\w^2)$, where $\w_{p, \HI}^\p$ is the dark plasma frequency  and $E_0^\p$ the binding energy associated with such ``dark hydrogen." We find that dark hydrogen's ability to weaken existing bounds on massive dark photons is suppressed compared to models of free mCPs, as considered previously in this section. To see this, note that at large $z$, when $\w = \w_0 (1+z) \gg E_0^\p$, $\meff^{\p \, 2} \to \w_{p, \HI}^{\p \, 2}$ and thus the effect of such particles on the effective dark photon mass is no different than that of freely ionized mCPs. However, at much later times when $\w \ll E_0^\p$, $\meff^{\p \, 2} \to  - \w_{p, \HI}^{\p \, 2} \,  (\w/ E _0^{\p})^2$. Hence, the typical effect of dark bound states is to induce a transition rather than forbid one. As a result, such models are more severally bounded than free mCPs. We leave a more detailed investigation to future work.

\section{Conclusions}
\label{sec:conclusion}
In this paper, we studied the prospects of detecting massless dark photons through a population of cosmic relics charged under the corresponding U(1) gauge symmetry. Such relics can make up a small fraction of the universe's energy budget but still substantially alter the massless dark photon dispersion relation by inducing a redshift-dependent dark plasma mass. If during the cosmic history, this effective mass is comparable to the Standard Model photon plasma mass, then a kinetic mixing between the sectors will induce $\g \leftrightarrow \Ap$ interconversion. Most importantly, photons from the cosmic microwave background can convert to dark photons, introducing detectable spectral distortions in their spectrum. 

We began by presenting a pedagogical introduction to the millicharged particle-induced $\gamma \leftrightarrow A'$ resonant interconversion and the resulting distortion of the CMB away from a blackbody spectrum. We then applied the results to both dark matter and dark radiation. For a dark matter subcomponent generated from thermal freeze-out, we showed that the bounds on millicharged particles from CMB spectral distortions are complimentary to stellar and cosmological limits. Furthermore, for heavier dark matter, spectral distortions can close parameter space corresponding to the ``overburden region" where dark matter is strongly interacting in the $(1-100)\ \gev$ range and direct detection bounds are substantially weakened. For millicharged dark radiation, we find that FIRAS is sensitive to tiny energy densities, well below current cosmological limits on additional neutrino degrees of freedom. Furthermore, a future satellite experiment similar to the proposed PIXIE mission would be capable of probing the abundance of millicharged particles generated from freeze-in. In the final part of our paper, we discussed how bounds on massive dark photons are altered in the presence of cosmic relics, finding modifications to previously computed bounds that assume the absence of a millicharged particle background. Our results stress the importance of terrestrial experiments as probes of dark photons, since they are less sensitive to the cosmic history of our Universe. 

We conclude here by commenting on several possible future developments. First, we note that in our analysis, we treated the universe as homogeneous. Inhomogeneities in the Standard Model plasma (as previously considered in Refs.~\cite{Caputo:2020bdy,Caputo:2020rnx} for massive dark photons) or in the dark plasma can also induce resonant oscillations during low redshifts. This motivates future work to model both the electron and millicharged particle density fluctuations and their impact on CMB spectral distortions. Second, in our study of dark radiation, we considered two possible extremes for the phase space distribution: a thermal phase space peaked at $p_\chi \sim T_\x$ and a freeze-in phase space peaked at $p_\chi \sim T_\g$. The $\g \to \Ap$ transition probability depends sensitively on the millicharged particle phase space density and hence plays an important role in deriving the corresponding constraints. Incorporating the phase space evolution due to self-interactions requires solving the full set of Boltzmann equations, which we leave to future work. Lastly, dark sector self-interactions can also suppress the dark plasma mass (as evident from \Eq{omega1body}), which in turn suppresses the likelihood of $\g \to \Ap$ oscillations. This effect is potentially important for ultralight millicharged particles~\cite{Bogorad:2021uew}, where the large phase space density may lead to Bose-enhanced processes. We leave a detailed study of this regime to future work.

\section*{Acknowledgments}
We thank Robert Lasenby, Di Liu, Hongwan Liu, Xingchen Xu, and Peizhi Du for useful discussions and Andrea Caputo for sharing preliminary results of Ref.~\cite{AndreaFuture}. This material is based upon work supported by the U.S. Department of Energy, Office of Science, National Quantum Information Science Research Centers, Superconducting Quantum Materials and Systems Center (SQMS) under contract number DE-AC02-07CH11359. Fermilab is operated by the Fermi Research Alliance, LLC under Contract DE-AC02-07CH11359 with the U.S. Department of Energy. JAD is supported in part by NSF CAREER grant PHY-1915852.  JTR is supported by NSF grants PHY-1915409 and PHY-2210498.  JTR acknowledges hospitality from the Aspen Center for Physics, which is supported by NSF grant PHY-1607611.
XG is supported by James Arthur Graduate Associate (JAGA) Fellowship.    This work was supported in part through the NYU IT High Performance Computing resources, services, and staff expertise.

\appendix

\section{Fluid Derivation of $\g$ and $\Ap$ Dispersion Relations}
\label{app:fluid}

In this appendix, we derive the equations of motion for the SM and dark photon fields, as shown in \Eq{disp1body}. Here, for simplicity we take the mCPs to be non-relativistic. The starting point of our calculation is thus to write down the coupled equations of motion for the SM plasma (consisting of free electrons, protons, and neutral hydrogen) and the mCP plasma (consisting of dark sector particles $\x^\pm$ that are oppositely charged under the dark photon) in the early universe. Since the proton-to-electron mass ratio is very large, we only include the free and bound electron contributions to the SM plasma. In the fluid approximation, the equation of motion for the free electrons is given by the non-relativistic limit of the Euler equation~\cite{Landau1987Fluid,Dubovsky:2015cca}
\be
\label{eq:efEOM}
\partial_t \vv_e \simeq - \frac{e}{m_e} \, \E - \Gamma \, \vv_e
~,
\ee
where $\vv_e$ is the bulk velocity of the free electron fluid, $\E$ is a background electric field, and $\Gamma$ is the collisional drag-rate between electrons and heavier SM species (i.e, the inverse timescale for an electron to exchange an $\order{1}$ fraction of its momentum).\footnote{In writing down \Eq{efEOM}, we have neglected subleading terms that are higher order in the electron velocity, such as those arising from magnetic fields.} In the Lorentz oscillator model, the electrons $e_b$ bound in neutral hydrogen are instead treated as a harmonic oscillator driven by the background electric field and with a restoring force governed by the characteristic binding energy of hydrogen $E_0 \simeq \alpha^2 m_e / 2$ (not to be confused with the electric field),
\be
\label{eq:ebEOM}
\ddot{\xv}_{e_b} + E_0^2 \, \xv_{e_b} \simeq - \, \frac{e}{m_e} \, \E
~,
\ee
where $\xv_{e_b}$ is the position of a bound electron. For the dark fluid, we take the mCPs to be free ionized particles since in the parameter space of interest, the mCP self-coupling $e^\p$ is not sufficiently large to efficiently form dark bound states.  Working in the diagonalized basis in which $j_\mu^\p$ couples to the linear combination $\Ap_\mu + \eps A_\mu$, as in the second equality of \Eq{lagrangian}, the bulk velocities of the $\x^\pm$ fluids evolve analogous to \Eq{efEOM} as
\be
\label{eq:xEOM1}
\partial_t \vv_\pm \simeq \pm \frac{e^\p}{m_\x} \, (\E^\p + \eps \, \E) - \Gamma^\p \, (\vv_\pm - \vv_\mp)
~,
\ee
where $\Gamma^\p$ is the collisional drag-rate between the $\x^+$ and $\x^-$ fluids. Taking the difference of the $\vv_+$ and $\vv_-$ components of \Eq{xEOM1} and defining the relative mCP bulk velocity $\vv_\x \equiv \vv_+ - \vv_-$, we thus have
\be
\label{eq:xEOM2}
\partial_t \vv_\x \simeq \frac{2 e^\p}{m_\x} \, (\E^\p + \eps \, \E) - 2 \Gamma^\p \, \vv_\x
~.
\ee
Finally, we supplement the above fluid equations with the electromagnetic equations of motion.  In Lorenz gauge ($\partial_\mu A^\mu = \partial_\mu A^{\p \, \mu} = 0$) and once again working in the diagonal basis in which the photon field $A^\mu$ couples to both the visible and dark currents, these are given by
\be
\label{eq:Maxwell1}
\partial^2 A_\mu = j_\mu + \eps \, j^\p_\mu
~~,~~
\partial^2 \Ap_\mu = j^\p_\mu 
~.
\ee

In order to proceed, we solve Eqs.~(\ref{eq:efEOM}), (\ref{eq:ebEOM}), and (\ref{eq:xEOM2}) for the velocities $\vv_e$, $\vv_{e_b}$, and $\vv_\x$, respectively, and use these to evaluate the SM and dark current densities $j$ and $j^{\p}$ in \Eq{Maxwell1}. This leaves a coupled set of equations for $A$ and $A^\p$, which can be diagonalized to give the dispersion relations for the SM-like and dark-like photon fields. This is simplest to do in Fourier space\footnote{We adopt the convention that the Fourier transform $f(k, \w)$ of a function $f(x,t)$ satisfies $f(x,t) \propto \int d^3 \kv \, d \w ~ e^{i (\kv \cdot \xv - \w t)} \, f(k,\w)$.} in which case we find that the free electron, bound electron, and mCP bulk velocities are
\begin{align}
\label{eq:velocities}
\tilde{\vv}_e \simeq - \frac{i e \,  \tilde{\E}}{m_e \, \w \, (1 + i \Gamma/\w)}
~~,~~
\tilde{\vv}_{e_b} \simeq \frac{i \w \, e \, \tilde{\E}}{m_e \, (E_0^2 - \w^2)} 
~~,~~
\tilde{\vv}_\x \simeq \frac{2 i e^\p \, (\tilde{\E}^\p + \eps \tilde{\E})}{m_\x \, \w \, (1 + 2 i \Gamma^\p/\w)}
~.
\end{align}
In \Eq{velocities} and in the remainder of this section, all quantities are evaluated in Fourier space and assumed to be functions of $\kv$ and $\w$ (Fourier transforms are denoted by a tilde).\footnote{Note that $\vv_e$ and $\vv_{e_b}$ have the opposite sign for $\Gamma \ll \w \ll E_0$; as a result, free and bound electrons typically have opposite effects on the dispersion relation of the SM photon.} \Eq{velocities} can then be used to determined the spatial part of the currents $\jv = - e \, ( n_e \vv_e + n_\HI \vv_{e_b})$ and $\jv^\p = e^\p n_\x \vv_\x / 2$, where $n_e$, $n_{\HI}$, and $n_\x = n_{\x^+} + n_{\x^-}$ are the free electron, bound electron (approximately equal to the neutral hydrogen $\HI$ density), and total mCP number densities. In particular, we find
\be
\label{eq:Euler3}
\tilde{\jv} \simeq \frac{i }{\w} \, \meff^2 \, \tilde{\E}
~~,~~
\tilde{\jv}^\p \simeq \frac{i}{\w} \, \meff^{\p \, 2} \, (\tilde{\E}^\p + \eps \tilde{\E})
~,
\ee
where we have defined the effective plasma masses (see the dispersion relations in \Eq{disp1} below)
\be
\label{eq:omega1}
\meff^2 \equiv \frac{\w_{p, e}^2}{1 + i \Gamma/\w} - \frac{\w_{p, \HI}^2}{E_0^2/\w^2 - 1}
\quad,\quad
\meff^{\p \, 2} \equiv \frac{\w_p^{\p \, 2}}{1 + 2 i \Gamma^\p/\w} 
~.
\ee
Above, $\w_{p,e}^2 = e^2 n_e / m_e$ and $\w_{p,\HI}^2 = e^2 n_\HI / m_e$ are the contributions to the SM photon plasma frequency from free electrons and electrons bound in neutral hydrogen, respectively, whereas $\w_p^{\p \, 2} = e^{\p \, 2} n_\x / m_\x$ is the mCP contribution to the dark photon plasma frequency. Note that in the limit that the free electron collisional rate is small and the hydrogen binding energy is large (i.e., $\Gamma \ll \w \ll E_0$), then $\meff^2 \simeq (1.4 \times 10^{-21} \ \eV^2) \, (n_p/\cm^{-3}) - (7.4 \times 10^{-24} \ \eV^2) \, (\omega/\eV)^2 (n_{\HI}/\cm^{-3})$.

Before substituting \Eq{Euler3} into the source terms of Maxwell's equations in \Eq{Maxwell1}, we wish to express all electromagnetic fields in terms of the electromagnetic potentials $A^\mu$ and $A^{\p \, \mu}$, including the electric fields that appear in \Eq{Euler3}. To do this, we decompose the vector potentials into transverse and longitudinal components, $\tilde{\A} = \tilde{\A}_T + (\kv / k) \, \tilde{A}_L$ with $\kv \cdot \tilde{\A}_T = 0$, which along with the Lorenz gauge condition yields $\tilde{A}^0 = (k / \w) ~ \tilde{A}_L$. This implies that the electric field $\E = - \grad A^0 - \partial_t \A$ is decomposed in terms of transverse and longitudinal components as
\be
\label{eq:decomp}
\tilde{\E} = i \w \Big[ \tilde{\A}_T + \Big( 1 - \frac{k^2}{\w^2} \Big) \, \frac{\kv}{k} \, \tilde{A}_L \Big]
~,
\ee
and similarly for the dark electric field $\E^\p$. Finally, using Eqs.~(\ref{eq:Euler3}) and (\ref{eq:decomp}) in the Fourier transform of \Eq{Maxwell1}, we find
\be\label{eq:disp1}
(\w^2 - k^2) \, 
\begin{pmatrix}
\tilde{\A}_T \\ \tilde{\A}_T^\p
\end{pmatrix}
\simeq 
\begin{pmatrix}
\meff^2 + \eps^2 \, \meff^{\p \, 2} & \eps \, \meff^{\p \, 2} \\
 \eps \, \meff^{\p \, 2}  & \meff^{\p \, 2}
\end{pmatrix}
\begin{pmatrix}
\tilde{\A}_T \\ \tilde{\A}_T^\p
\end{pmatrix}
~~,~~
\w^2 \, 
\begin{pmatrix}
\tilde{\A}_L \\ \tilde{\A}_L^\p
\end{pmatrix}
\simeq 
\begin{pmatrix}
\meff^2 + \eps^2 \, \meff^{\p \, 2} & \eps \, \meff^{\p \, 2} \\
 \eps \, \meff^{\p \, 2}  & \meff^{\p \, 2}
\end{pmatrix}
\begin{pmatrix}
\tilde{\A}_L \\ \tilde{\A}_L^\p
\end{pmatrix}
~,
\ee
which is the starting point shown in \Eq{disp1body} of the calculation in \Sec{transitions}. Note that while this calculation was performed for non-relativistic mCPs, it can be simply generalized to relativistic mCPs provided that self-interactions are negligible. This corresponds to setting $\Gamma^\p = 0$ and modifying $\meff^{\p \, 2} \simeq \w_p^{\p \, 2} \to \meff^{\p \, 2} \simeq (3/2) \, \w_p^{\p \, 2}$~\cite{Braaten:1993jw,Raffelt:1996wa,Bellac:2011kqa}, as in \Eq{omega1body}.

\section{Alternative Derivations of the Resonant Transition Rate}
\label{app:resonant}

In \Sec{transitions} (and \App{fluid}) we worked in frequency-space to derive the form for the resonantly enhanced $\g \to \Ap$ transition rate, commonly referred to as the Landau-Zener formula~\cite{landau1932theorie,Zener:1932ws}. In this section, we provided two alternative derivations of this result, using either a quantum mechanical formulation (that works explicitly in terms of time instead of frequency), or the Boltzmann equation. For simplicity, we focus on the transition between transverse modes.

\subsection{Schr\"{o}dinger Equation}

We begin with the traditional time domain method which was first developed by Landau and Zener~\cite{landau1932theorie, Zener:1932ws}, and later used within the context of neutrino physics~\cite{Parke:1986jy, Kuo:1989qe} (see also the discussion in Ref.~\cite{Caputo:2020rnx}). Working in the ultrarelativistic and collisionless limit, the dispersion relation of \Eq{disp1body} corresponds to (up to an arbitrary phase) the following Schr\"{o}dinger-type time-dependent equation
\be
\label{eq:Schrodinger_Eq}
i \partial_t 
\begin{pmatrix}
\A_T \\
\A'_T
\end{pmatrix} \simeq 
\begin{pmatrix}
\xi & \eta \\
\eta & - \xi 
\end{pmatrix}
\begin{pmatrix}
\A_T \\
\A'_T 
\end{pmatrix}
~,
\ee
where $\xi = (\meff - \meff^\p)/4\w$ and $\eta = \eps \, \meff/2\w$ (we remind the reader that we have shortened our notation such that only the real part is included in $\meff^{2}$ and $\meff^{\p \, 2}$). Now, expanding $\A_T$ and $\A_T'$ in terms of the instantaneous basis of the $\order{\eps^0}$ Hamiltonian, we have
\be
\label{eq:AAp_Expand}
\A_T =  c_{\g}(t)\,e^{ -i\int_{-\infty}^t dt' \, \xi(t')}
~,~
\A_T' = c_{\Ap}(t)\,e^{ i\int_{-\infty}^t dt' \, \xi(t')}
~.
\ee
To determine the time-dependent coefficients $c_\g$ and $c_{\Ap}$, we substitute \Eq{AAp_Expand} into \Eq{Schrodinger_Eq}, which yields the following differential equations,
\be
\label{eq:c_EOM}
i \partial_t c_\g = \eta \, c_{\Ap} \, e^{2i \int_{-\infty}^t d t' \xi(t')}
~,~
i \partial_t c_{\Ap} = \eta \, c_{\gamma} \, e^{-2i \int^t_{-\infty} dt' \xi(t')}
~.
\ee
Since \Eq{c_EOM} is symmetric under $c_\g \leftrightarrow c_{\Ap}$, it follows that conversion probability satisfies $P_{\gamma \rightarrow \Ap} = P_{\Ap \rightarrow \gamma}$. Assuming an initially negligible dark photon density, $\gamma \to \Ap$ dominates over the inverse process, corresponding to the initial condition $c_{\gamma}(-\infty)=1$, $c_{\Ap}(-\infty)=0$. For $\eps \ll 1$, $c_\g (t)\simeq 1$ and the $\g \to \Ap$ transition probability at $t = \infty$ (to leading order in $\eps$) is approximately
\be
P_{\gamma \rightarrow A'} \simeq |c_{\Ap} (\infty)|^2 \simeq \abs{ \eta \int^{+\infty}_{-\infty} dt' e^{-2i \int^{t'}_{-\infty} dt'' \xi(t'')}  }^2
~.
\ee
This integral can be evaluated using the saddle point approximation, which gives
\be
\label{eq:P_AToAp_1stOrder}
P_{\g \to \Ap} \simeq  \sum_{t_\text{res}}\pi \, \eta^2 \, \abs{\frac{d \xi}{dt}}^{-1} \bigg|_{t_\text{res}} 
~,
\ee
where the sum includes all times $t_\text{res}$ at which $\xi(t_\text{res}) = 0$. Using the definitions of $\eta$ and $\xi$, we find that \Eq{P_AToAp_1stOrder} gives $P_{\g \to \Ap}$ agrees with  $- \Delta f_\g / f_\g$, where  $\Delta f_\g / f_\g$ is given by integrating \Eq{rate2} over time.

In order to calculate $P_{\g \to \Ap}$ to the all orders in $\epsilon$, we utilize the Dykhne-Davis-Pechukas (DDP) method~\cite{dykhne1962adiabatic, davis1976nonadiabatic, berry1990histories, vitanov1999nonlinear, schilling2006nonadiabatic}. This gives 
\be
\label{eq:DDP1}
P_{\gamma \rightarrow A'} = 1 - \prod_{t_\text{res}}e^{ - \frac{1}{\omega} \Im \int^{t_c}_0 dt ~ \left( \Pi_+ - \Pi_- \right)}
~,
\ee
where 
\be
\label{eq:Pi_plusminus}
\Pi_\pm = \frac{\meff^2 + \meff^{\p \, 2}}{2} \pm \frac{1}{2} \, \sqrt{ \bigg(\frac{d \left( \meff^2 - \meff^{\p \, 2} \right)}{dt}\bigg|_{t_\text{res}} \, \bigg)^2 \, t^2 + 4 \, \eps^2 \,  \meff^4 }
\ee
and
\be
t_c = 2 i \, \eps \, \meff^2 \, \abs{\frac{d \left( \meff^2 - \meff^{\p \, 2} \right)}{dt}}^{-1} \,  \Bigg|_{t_\text{res}}
~.
\ee
Performing the time integral in \Eq{DDP1} from $t=0$ to $t=t_c$, we find that the transition probability to all orders in $\eps$ is
\be
\label{eq:P_AToAp_AllOrder}
P_{\g \to \Ap} \simeq 1 - \exp \left( \sum_{t_\text{res}}\pi \, \eta^2 \, \abs{\frac{d \xi}{dt}}^{-1} \bigg|_{t_\text{res}}  \right)
~.
\ee
The probability for $\g \to \g$ is given by $P_{\g \to \g} = 1 - P_{\g \rightarrow \Ap}$.

\begin{figure}[t]
\includegraphics[width=10cm]{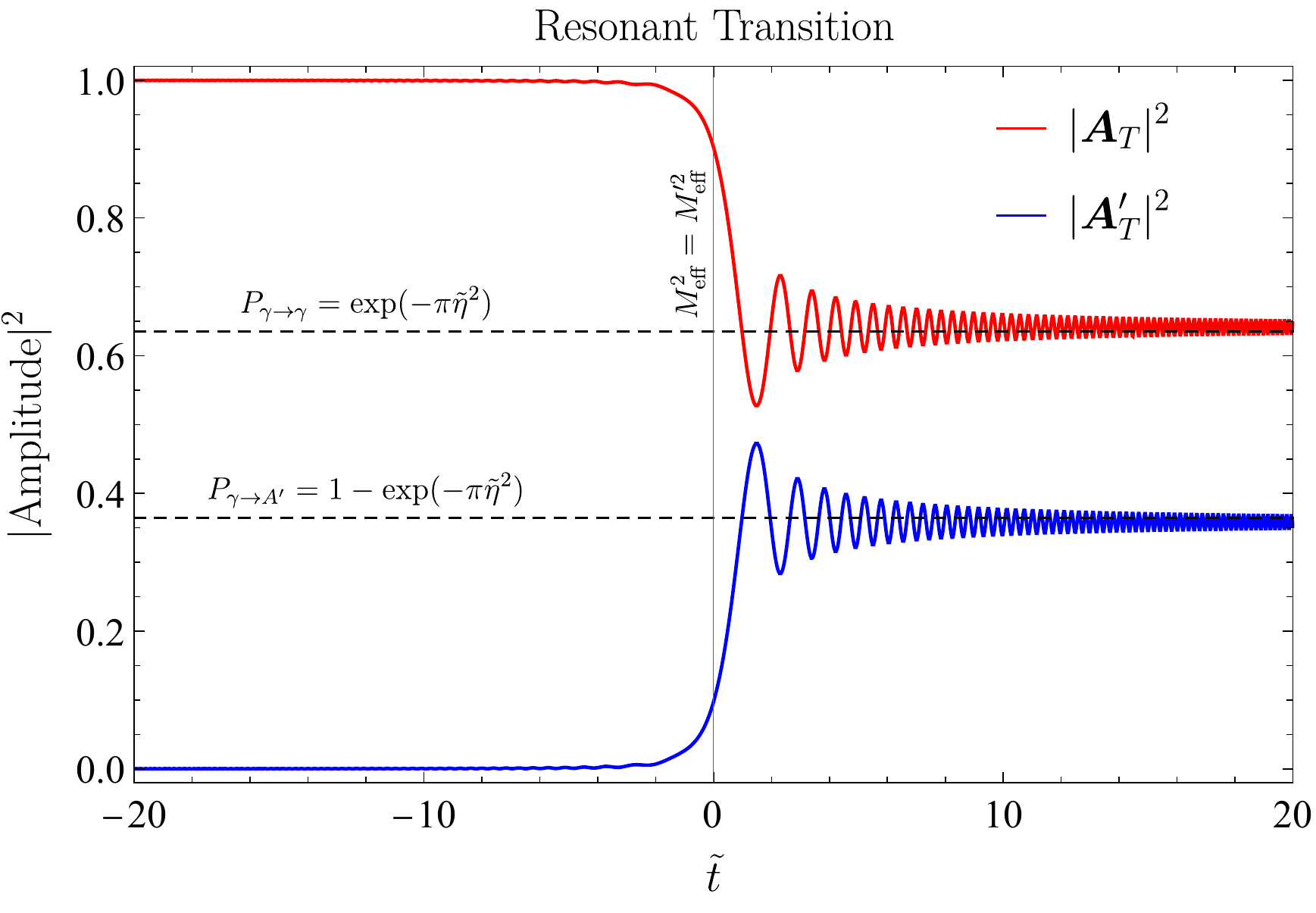}
\centering
\caption{The time evolution (in terms of the dimensionless time variable $\tilde{t}$) of the transverse field components $\A_T$ and $\A_T^\p$, as described by the two-level system of \Eq{Schrodinger_Eq}. Numerical solutions of \Eq{Schrodinger_Eq} are shown as red and blue lines, for the SM-like and dark-like photon states, respectively. Along the dashed gray lines, we show the late-time semi-analytic estimate, provided by \Eq{P_AToAp_AllOrder}. Resonant enhancement occurs at $\tilde{t} = 0$, corresponding to $\meff \simeq \meff^\p$ near the vertical gray line.}
\label{fig:num_resonant}
\end{figure}

In \Fig{num_resonant}, we compare the numerical solution of  \Eq{Schrodinger_Eq} ($|c_\g|^2$ in red and $|c_{\Ap}|^2$ in blue) to the semi-analytic approximation of \Eq{P_AToAp_AllOrder} (dashed gray lines). In the figure, we have defined the dimensionless quantities $\tilde{t} \equiv t \, \abs{ d \xi(t_\text{res}) / d t}^{-1/2}$ and $\tilde{\eta} \equiv \eta \, \abs{d \xi (t_\text{res}) / d t}^{-1/2}$. The resonantly enhanced transitions occurs when $\meff^2 \simeq \meff^{\p \, 2}$, denoted as $\tilde{t} \simeq 0$ near the vertical gray line. Before this time, $\abs{c_\g} \simeq 1$ and $\abs{c_{\Ap}} \simeq 0$. Near the resonance, photons convert to the dark photons. After the resonance, $c_\g$ and $c_{A^\p}$ oscillate around the asymptotic value given in \Eq{P_AToAp_AllOrder}. 

%

\subsection{Boltzmann Equation}

In this section we provide an alternative derivation of the $\g \to \Ap$ transition rate, utilizing an explicit calculation involving the Boltzmann equation. The modification of the SM photon phase space density $f_{\g}$ due to $\g \rightarrow \Ap$ is described by 
\be
\label{eq:Int_Boltz_Eq_AToAp}
\frac{1}{a^3}\frac{\partial \left( a^3 f_\gamma  \right)}{\partial t}\simeq - f_\gamma \, \Gamma_{\gamma \rightarrow \Ap},
\ee
where $a$ is the scale factor,
\be
\label{eq:Gamma_AToAp}
\Gamma_{\gamma \rightarrow \Ap} 
= \frac{g_{\Ap}}{2 \omega_\g} \int \frac{ d^3 k_{\Ap} }{2 \omega_{\Ap} \, (2 \pi)^3} \left( 2 \pi \right)^4 \delta^{(4)}\left(k_\g - k_{\Ap}\right) \abs{\mathcal{M}_{\gamma \rightarrow \Ap}}^2
~,
\ee
$\w_i$, $k_i$, and $f_i$ are the energy, momentum, and phase space density of species $i$, $|\mathcal{M}_{\Ap \to \g}|^2$ is the squared matrix element for $\Ap \to \g$ averaged over initial and final spin, $g_{\Ap} = 2$ is the number of internal $\Ap$ spin degrees of freedom, and we have taken $f_{\Ap} \simeq 0 $~\cite{Kolb:1990vq}. 

From Ref.~\cite{An:2020jmf}, we have that $\mathcal{M}_{\gamma \rightarrow \Ap} = \epsilon \, \meff^2 \, \epsilon_{\mu}^i(k_\g) \,  \epsilon^j_{\mu}(k_{\Ap})$, where $\eps(k)^i$ are polarization vectors. Averaging over the SM and dark photon polarizations, this gives $\abs{\mathcal{M}_{\gamma \rightarrow \Ap}}^2 \rightarrow \epsilon^2 \meff^4/2$, such that \Eq{Int_Boltz_Eq_AToAp} can be rewritten as 
\be
\frac{1}{a^3}\frac{\partial \left( a^3 f_\gamma  \right)}{\partial t} \simeq - f_\gamma \frac{\pi \, \epsilon^2 \, \meff^4 }{ 2 \omega_\g^2} \, \delta\left(\omega_\g - \omega_{\Ap}\right)
~,
\ee
where $\w_\g \simeq k_\g + \meff^2 / 2 k_\g$, $\w_{\Ap} \simeq k_{\Ap} + \meff^{\p \, 2} / 2 k_{\Ap}$, and $k_\g = k_{\Ap}$ by energy-momentum conservation. Hence, we finally arrive at
\be
\frac{1}{a^3}\frac{\partial \left( a^3 f_\gamma  \right)}{\partial t} \simeq - \frac{\pi \, \epsilon^2 \,  f_\gamma}{\omega_\g} \, \meff^4  \, \delta\left(\meff^2 - \meff^{\p \, 2} \right)
~,
\ee
which agrees with \Eq{rate2} in the limit that we can ignore variations to the scale factor during the moment of resonance.

\section{Freeze-in of mCDR}
\label{app:freezein}

In Sec. \ref{sec:DRlimit}, we discussed signals arising from massless dark photons in the presence of mCDR\@. In particular, we discussed the minimal cosmology in which the mCDR energy density is frozen in from plasmon decay, $\g^* \to \x^+ \x^-$. In this section, we provide a detailed derivation of the number density, $n_\x$, energy density, $\rho_\x$, and resulting effective dark photon mass, $\meff^\p$, produced from mCDR freeze-in. In doing so, we follow the calculations detailed previously in Ref.~\cite{Dvorkin:2019zdi,  Chu:2011be, Hambye:2019dwd}. 

Our starting point is the Boltzmann equation for massless mCPs produced from plasmon decay  $\g^* \to \x^+ \x^-$~\cite{Kolb:1990vq},
\be
\label{eq:Full_BoltzEq}
\frac{ \partial f_{\chi^+}}{ \partial t } - H \, p_{\chi^+} \, \frac{ \partial f_{\chi^+}}{ \partial p_{\chi^+}} \simeq \frac{g_\g \, g_{\x^-}}{2 p_{\chi^+}} \, \int \frac{\dbar^3 k_\g}{2 \omega_\g} ~ \frac{\dbar^3 p_{\chi^-}}{2p_{\chi^-}}  ~ f_\gamma ~ (2\pi)^4 \delta^{(4)}\left(k_\g - p_{\chi^+} -p_{\chi^-} \right) ~ \overline{\left| \mathcal{M}\right|^2}
~,
\ee
where $\overline{\left| \mathcal{M}\right|^2}$ is the matrix element for plasmon decay (averaged over initial and final states),  $\dbar^3 p \equiv d^3 p/(2\pi)^3$, and we have neglected the inverse process $\chi^+ \chi^- \rightarrow \gamma^*$. This can conveniently rewritten in terms of the plasmon decay rate $\Gamma_{\g^* \to \x^+ \x^-}$ as
\be
\label{eq:Full_BoltzEq2}
\frac{ \partial f_{\chi^+}}{ \partial t } - H \, p_{\chi^+} \, \frac{ \partial f_{\chi^+} }{ \partial p_{\chi^+}} \simeq \frac{n_\g}{g_{\x^+}} \, \bigg\langle \frac{d \Gamma_{\g^* \to \x^+ \x^-}}{\dbar^3 p_{\x^+}} \bigg\rangle
~,
\ee
where $n_\gamma \simeq (g_\g \, \zeta(3) / \pi^2) \, T^3$ is the plasmon number density and the brackets corresponds to a thermal average over the plasmon phase space. For the remainder of this section, we focus exclusively on the decay of transverse plasmons ($g_\g = 2$), since the longitudinal contribution is subleading~\cite{Dvorkin:2019zdi}. Furthermore, for fermionic mCDR we set $g_{\x^+} = g_{\x^-} = 2$.

\subsection{mCP Number Density}

Integrating \Eq{Full_BoltzEq2} over the mCP phase space and defining the total mCP number density as $n_\x = n_{\x^+} + n_{\x^-}$, we have
\be
\label{eq:nchi1}
\dot{n}_{\chi} + 3 H n_{\chi} = 2n_\gamma \left\langle \Gamma_{\g^* \to \x^+ \x^-}  \right\rangle
~,
\ee
where from Ref.~\cite{Dvorkin:2019zdi} the thermally-averaged decay rate is approximately 
\be
\left\langle \Gamma_{\gamma^* \rightarrow \x^+ \x^-}  \right\rangle
 \simeq \frac{\pi \, (e q_\chi)^2}{144 \, \zeta(3)} \, \frac{\meff^2}{T_\g}
 ~.
\ee
Assuming an initially negligible abundance of mCDR, \Eq{nchi1} can then be solved in terms of the comoving mCP number density (yield) $Y_\chi = n_\chi/s_\gamma$ at late times,
\be
Y_{\chi} \simeq \int^{\infty}_{0} dT_\g ~ \frac{2 n_\gamma}{s_\gamma H \, T_\g} ~ \left\langle\Gamma_{\g^* \to \x^+ \x^-}\right\rangle
\simeq 3.3 \times 10^{-12} \, \left(\frac{q_\chi}{10^{-14}}\right)^2
~.
\ee
%

\subsection{mCP Energy Density}

Multiplying \Eq{Full_BoltzEq2} by $p_\x$ and then integrating over the mCP phase space yields the following equation for the total mCP energy density
\be
\label{eq:rhochi1}
\dot{\rho}_{\chi} + 4 H \rho_{\chi} = n_\gamma \left\langle \omega_\g \, \Gamma_{\gamma^* \rightarrow \x^+ \x^-}\right\rangle, 
\ee
where 
\be
\left\langle \omega_\g \, \Gamma_{\g^* \to \x^+ \x^-} \right\rangle 
 \simeq \frac{(e q_\chi)^2}{12 \pi} \, \meff^2
 ~.
\ee
Analogous to $Y_\chi$, we define the quantity $A_{\chi} \equiv \rho_{\chi}/s_\g^{4/3}$. Similar to the previous subsection, \Eq{rhochi1} can then be solved to determine $A_\x$ at late times
\be
\label{eq:A_XXbar}
A_{\chi} = \int^{\infty}_{0}  dT ~ \frac{n_\gamma}{s_\g^{4/3}  H \, T_\g}  \, \left\langle \omega_\g \, \Gamma_{\gamma^* \rightarrow \x^+ \x^-} \right\rangle
~.
\ee
We evaluate \Eq{A_XXbar}  numerically in order to determine the additional number of effective neutrino degrees of freedom in mCDR\@. This gives
\be
\label{eq:DelNeff_BoltzEq}
\Delta N_\eff \simeq 2 \times 10^{-11} \left(\frac{q_\chi}{10^{-14}}\right)^2
~.
\ee
%

\subsection{Dark Plasmon Mass} 

From \Eq{mApDR}, the effective dark photon mass arising from mCDR is
\be
\meff^{\p \, 2}   \simeq \frac{8 \alpha^\p}{\pi} \int^{\infty}_0 d p_\chi ~ p_\chi \, f_\chi
~.
\ee
Multiplying \Eq{Full_BoltzEq2} by $1/p_\x$ and then integrating over the mCP phase space gives
\be
\label{eq:Meffeq1}
\frac{\partial \meff^{\p \, 2}}{\partial t} + 2 H \, \meff^{\p \, 2}  = 16 \pi \alpha^\p \, n_\gamma \, \left\langle \frac{\Gamma_{\gamma^* \rightarrow \x^+ \x^-}}{\omega_\g} \right\rangle
~,
\ee
where 
\be
\label{eq:Gamma_Over_omegaT}
\left\langle \frac{\Gamma_{\gamma^* \rightarrow \x^+ \x^-}}{\omega_\g} \right\rangle \simeq \frac{(e q_\x)^2}{24 \pi \, \zeta(3)} \frac{\meff^2}{ T_\g^2} \, \log{\bigg(\frac{T_\g}{\meff}\bigg)}
~.
\ee
Solving \Eq{Meffeq1} for $B_{\chi} \equiv \meff^{\p \, 2}/s_\g^{2/3}$ yields
\be
B_{\chi} \simeq 16 \pi \alpha^\p \int^{\infty}_{0} d T_\g ~ \frac{n_\gamma}{s^{2/3}_\gamma H \, T_\g} \left\langle \frac{\Gamma_{\gamma^* \rightarrow \x^+ \x^-}}{\omega_\g} \right\rangle
~.
\ee
The above expression can be evaluated numerically, which gives
\be
\label{eq:mAp_FreezeIn_BoltzEq}
\meff^\p \simeq 3 \times 10^{-17} \ \eV \times \bigg(\frac{q_\chi}{10^{-14}}\bigg)^2 \bigg(\frac{10^{-7}}{\epsilon}\bigg) \, \left(1+z\right)
~.
\ee

\bibliographystyle{apsrev4-1}
\bibliography{references}

\end{document}